\documentclass[manuscript, nonacm]{acmart}

\AtBeginDocument{%
  }

\setcopyright{acmlicensed}
\copyrightyear{2025}
\acmYear{2025}
\acmDOI{XXXXXXX.XXXXXXX}
\acmConference[OzCHI '25]{the 37th Australian Conference on Human-Computer Interaction}{2025}{Sydney, Australia}
\acmISBN{978-1-4503-XXXX-X/2025/12} 

\usepackage{booktabs}       
\usepackage{array}          
\usepackage{makecell}       
\usepackage{longtable}      
\usepackage{threeparttable} 
\usepackage{tabularx}       

\usepackage{tikz}           
\usetikzlibrary{shapes,arrows,positioning,calc,arrows.meta,shapes.geometric}
\usepackage{adjustbox}      

\usepackage{listings}       

\usepackage{xcolor}         

\usepackage{pifont} 

\usepackage[font=small,labelfont=bf]{caption}

\begin{document}
\captionsetup[table]{font=normalsize, labelfont=bf}
\captionsetup[figure]{font=normalsize, labelfont=bf}

\thanks{This is the author-prepared version of a paper accepted to OZCHI 2025. The official version will be published by ACM.}

\begin{center}
\fbox{
\parbox{0.96\linewidth}{
\textbf{Please cite as follows:}  
Mahmud, R., Wu, Y., Bin Sawad, A., Berkovsky, S., Prasad, M., and Kocaballi, A. B. (2025). \textit{Evaluating User Experience in Conversational Recommender Systems: A Systematic Review Across Classical and LLM-Powered Approaches}. In \textit{Proceedings of the Australian Human–Computer Interaction Conference (OZCHI 2025)}. ACM.
}
}
\end{center}
\vspace{1em}

\title [CRS UX: A Systematic Review] {Evaluating User Experience in Conversational Recommender Systems: A Systematic Review Across Classical and LLM-Powered Approaches}




\author{Raj Mahmud}
\email{raj.mahmud@uts.edu.au}
\affiliation{%
  \institution{School of Computer Science, University of Technology Sydney}
  \city{Sydney}
  \state{NSW}
  \country{Australia}
}

\author{Yufeng Wu}
\affiliation{%
  \institution{School of Computer Science, University of Technology Sydney
  \city{Sydney}
  \country{Australia}}
  }

\author{Abdullah Bin Sawad}
\affiliation{%
  \institution{The Applied College, King Abdulaziz University}
  \city{Jeddah}
  \country{Saudi Arabia}
}

\author{Shlomo Berkovsky}
\affiliation{%
 \institution{Australian Institute of Health Innovation, Macquarie University}
 \city{Sydney}
 \state{NSW}
 \country{Australia}}

\author{Mukesh Prasad}
\affiliation{%
  \institution{School of Computer Science, University of Technology Sydney}
  \city{Sydney}
  \state{NSW}
  \country{Australia}}

\author{A. Baki Kocaballi}
\email{Baki.Kocaballi@uts.edu.au}
\affiliation{%
  \institution{School of Computer Science, University of Technology Sydney}
  \city{Sydney}
  \state{NSW}
  \country{Australia}
}

\renewcommand{\shortauthors}{Mahmud et al.}

\begin{abstract}
Conversational Recommender Systems (CRSs) are receiving growing research attention across domains, yet their user experience (UX) evaluation remains limited. Existing reviews largely overlook empirical UX studies, particularly in adaptive and large language model (LLM)-based CRSs. To address this gap, we conducted a systematic review following PRISMA guidelines, synthesising 23 empirical studies published between 2017 and 2025. We analysed how UX has been conceptualised, measured, and shaped by domain, adaptivity, and LLM. Our findings reveal persistent limitations: post hoc surveys dominate, turn-level affective UX constructs are rarely assessed, and adaptive behaviours are seldom linked to UX outcomes. LLM-based CRSs introduce further challenges, including epistemic opacity and verbosity, yet evaluations infrequently address these issues. We contribute a structured synthesis of UX metrics, a comparative analysis of adaptive and nonadaptive systems, and a forward-looking agenda for LLM-aware UX evaluation. These findings support the development of more transparent, engaging, and user-centred CRS evaluation practices.
\end{abstract}



\maketitle

\section{Introduction}

Conversational Recommender Systems (CRSs) are interactive systems that support users in discovering, refining, and selecting items through natural-language dialogue. Unlike traditional recommenders that generate static suggestions based on past behaviour or item similarity, CRSs enable multi-turn, dynamic conversations in which users can iteratively express preferences, explore alternatives, and resolve ambiguities. Early CRS designs combined handcrafted dialogue flows with content-based or collaborative filtering engines \cite{mccarthy2004dynamic}, often relying on rule-based interaction patterns and constrained slot-filling strategies \cite{jannach2021survey}. Over time, the field has evolved to incorporate adaptive dialogue policies, mixed-initiative capabilities, and user-modelling techniques \cite{zhang2024navigating}. The recent surge of large language models (LLMs), such as GPT-4, has further transformed the design of CRS, enabling open-domain dialogue generation and more expressive, human-like conversational styles \cite{deldjoo2404review, gao2021advances}. However, these advances have also introduced new challenges in UX design, system transparency, and evaluation methodology.

As CRSs are deployed across domains including e-commerce, media, travel, and health \cite{jannach2021survey}, understanding how users perceive, interact with, and respond to these systems has become increasingly important. Several recent surveys have comprehensively reviewed technical progress in conversational recommendation, covering models, architectures, and dialogue policies \cite{gao2021advances, li2023conversation, zaidi2024review, lei2020conversational}, but these works pay limited attention to UX as a distinct evaluative concern. They largely focus on system-level capabilities such as natural language understanding, response generation, and recommendation accuracy, rather than the experiential dimensions of CRS interaction or empirical UX evaluation practices. While a few recent reviews, such as Jannach et al.~\cite{jannach2023evaluating}, have begun to foreground UX concerns and examine evaluation practices from a user-centric perspective, these works remain primarily conceptual or narrow in scope. A systematic, empirical synthesis of how UX is defined, operationalised, and measured in CRS research is still lacking. In particular, prior work has not fully accounted for domain-specific variation, adaptivity mechanisms, or the emergent UX challenges posed by LLM-powered CRS. Although constructs such as satisfaction, trust, and perceived usefulness are frequently measured, there is little consensus on how to evaluate the temporal, affective, and behavioural dynamics that shape user perceptions during interaction \cite{luger2016like}. Generative CRS further complicate evaluation by introducing epistemic opacity, verbosity control issues, and novel interaction risks that standard UX instruments may not capture effectively \cite{deldjoo2404review, zhao2024recommender}.
To address these gaps, we present a systematic literature review (SLR) of CRS studies with an explicit focus on UX evaluation. We apply the PRISMA methodology to identify and synthesise 23 empirical studies published up to May 2025. Our analysis focuses on four key questions:

\begin{itemize}
    \item \textbf{RQ1:} What UX dimensions are evaluated in CRS research, and how do these dimensions vary across application domains?
    \item \textbf{RQ2:} What evaluation methods and tools are used to assess UX in CRS, and what are their affordances and limitations?
    \item \textbf{RQ3:} How do adaptivity and personalisation influence the evaluation and experience of CRS?
    \item \textbf{RQ4:} What are the UX design and evaluation challenges posed by LLM-powered CRS?
\end{itemize}

Our contributions are fourfold. First, we provide a structured synthesis of the UX dimensions assessed in CRS research, categorised by domain. Second, we analyse the methods, tools, and measurement frameworks used to evaluate UX, highlighting common practices and gaps. Third, we offer a comparative analysis of adaptive versus nonadaptive systems, identifying how design choices affect reported user experiences. Fourth, we examine the emerging class of LLM-powered CRS, revealing distinctive UX challenges and underexplored risks that future research must address.

The remainder of this paper is organised as follows: Section~\ref{sec:related-work} reviews related work on CRS design and evaluation. Section~\ref{sec:methodology} outlines our review methodology. Section~\ref{sec:results} presents findings for RQ1--RQ4. Section~\ref{sec:discussion} synthesises implications and themes, and Section~\ref{sec:conclusion} concludes with reflections and future directions.

\section{Related Work}
\label{sec:related-work}

Research on CRS spans architectural design, user modelling, evaluation methodologies, and more recently, the integration of LLMs. This section groups prior work into four strands: architectural and evaluation-oriented surveys, empirical UX studies, LLM-focused CRS reviews, and a summary that positions our review within this landscape.

\subsection{Architectural and Evaluation-Oriented Surveys}
Several surveys have mapped the system-level evolution of CRS, often highlighting dialogue strategies, recommendation models, and conversational policies. Gao et al.~\cite{gao2021advances} provide a comprehensive overview of architectural components, including natural language understanding, state tracking, and policy learning. Their taxonomy helps clarify technical bottlenecks in multi-turn CRS design but offers limited treatment of UX dimensions or user-centred evaluation frameworks. Similarly, Lei et al.~\cite{lei2020conversational} formalise CRS as a joint optimisation task over dialogue and recommendation quality. They outline evaluation strategies ranging from offline ranking metrics to online A/B testing but do not examine user affect, trust, or satisfaction metrics in depth.
Zaidi et al.~\cite{zaidi2024review} review a broad spectrum of CRS applications, such as fashion, education, and tourism, highlighting domain-specific implementations and dialogue approaches. Their focus remains descriptive, offering little insight into user perception or temporal engagement. Pramod and Bafna~\cite{pramod2022conversational} examine techniques and tools for CRS development, including ontology-driven recommendation, critiquing interfaces, and adoption barriers. While they emphasise explainability and interface design, their work lacks a methodological breakdown of evaluation protocols or user-centred metrics.
In contrast, Jannach et al.~\cite{jannach2023evaluating} develop a dedicated evaluation framework, distinguishing three modes: offline simulations, lab-based studies, and in-the-wild deployments. They critique the over-reliance on accuracy metrics and advocate for more interaction-focused and affective evaluations. However, their review is broad and does not systematically extract constructs, instruments, or domains from empirical CRS studies. Wang et al.~\cite{wang2023rethinking} complement this by arguing that static evaluation pipelines do not reflect the dynamic nature of user–CRS interaction. While their proposals for interactive and human-centred benchmarks are conceptually rich, they remain largely untested and omit UX operationalisation practices used in prior studies.
Li et al.~\cite{li2023conversation} advocate for a holistic view of CRS, integrating multi-modal signals, user goals, and dialogue context. Although they stress the value of personalisation and interactivity, their review falls short of discussing how these qualities are evaluated in practice. Lian et al.~\cite{recai2024lian} introduce RecAI, a modular toolkit for building LLM-enabled recommender systems. While their architecture includes modules for explanation and dialogue monitoring, their evaluation approach focuses on internal benchmarks, not on empirical UX instruments or user feedback.

\subsection{Empirical UX Studies}
Numerous empirical studies explore UX in CRS, though typically with narrow or domain-specific scope. Siro et al.~\cite{siro2023understanding} apply turn-level annotation to the ReDial dataset, identifying satisfaction-related features such as interest arousal and repetition. Their work is valuable for modelling micro-level interaction signals but does not capture longitudinal satisfaction or user preferences. Ma and Ziegler~\cite{ma2024investigating} investigate proactive decision aids in CRS, revealing that system-initiated suggestions improve perceived usefulness, though at the cost of increased cognitive load. Their study highlights the UX trade-offs introduced by initiative strategies but lacks broader construct coverage.
Thom et al.~\cite{thom2024nutria} develop a GPT-powered nutrition assistant and evaluate it using CSAT, CES, and NPS scores. While the study presents useful deployment insights, it does not explore behavioural metrics, emotional reactions, or adaptation to user preferences. Kraus et al.~\cite{kraus2024pilot} conduct a Wizard-of-Oz study of group travel CRS, showing that conversation-leading strategies better align with user expectations when group personalities converge. Yet their work does not assess adaptivity mechanisms or engagement sustainability.
These studies collectively point to the need for systematic evaluation of diverse UX constructs, ranging from trust and clarity to surprise and engagement, but no synthesis currently exists to integrate these metrics across domains or system types.

\subsection{LLM-Focused CRS Reviews}

The introduction of LLMs has prompted new concerns in CRS design and evaluation. Wu et al.~\cite{wu2024survey} categorise LLM applications in recommendation into generation, planning, and reasoning. They identify new risks, including hallucination, response verbosity, and misalignment, but do not detail how such issues affect UX or how they are evaluated. Li et al.~\cite{li2024generative} similarly position LLM-based recommenders as generative pipelines but focus on system capabilities and challenges rather than user-facing evaluation methods.
Deldjoo et al.~\cite{deldjoo2404review} propose a conceptual taxonomy of evaluation risks introduced by generative models, distinguishing between exacerbated challenges (e.g., verbosity, latency) and novel ones (e.g., prompt brittleness, privacy leakage). While this framework is theoretically valuable, it is not grounded in empirical user studies. Huang et al.~\cite{huang2025agentic} envision agentic CRS that proactively support user goals, manage dialogue flow, and adapt to long-term preferences. Their vision points to new UX affordances but remains speculative in the absence of empirical validation.
Lian et al.~\cite{recai2024lian}, Hou et al.~\cite{hou2024zeroshot}, and others also explore how LLMs can function as zero-shot rankers or preference matchers, but these works focus on benchmark accuracy and fail to engage with experiential metrics such as trust, control, or expectation calibration. Across these reviews, the experiential implications of LLM deployment are acknowledged, yet the methods to evaluate such experiences remain vague or absent.

\subsection{Summary of Gaps and Our Contribution}

Despite significant growth in CRS and generative recommender literature, no prior review offers a structured synthesis of UX evaluation constructs, instruments, and methods across empirical studies. Existing surveys primarily focus on architectural taxonomies, dialogue strategies, or conceptual critiques. Even when UX is discussed, as in Jannach et al.~\cite{jannach2023evaluating} or Deldjoo et al.~\cite{deldjoo2404review}, coverage remains abstract, anecdotal, or limited to high-level dimensions. Critically, established UX instruments such as the User Experience Questionnaire (UEQ), System Usability Scale (SUS), and Recommender Quality Evaluation (ResQue) framework are seldom compared across studies. This lack of comparative analysis obscures how different tools capture affective, cognitive, or behavioural dimensions, and whether they are appropriate for CRS-specific evaluation across contexts. Moreover, domain-adaptive evaluation designs, those that account for setting-specific needs in domains like wellbeing, travel, or education, remain underexplored. This fragmentation hinders the development of cumulative knowledge about UX in CRS and limits reproducibility across systems and contexts.
In contrast, this review applies a PRISMA-guided methodology to systematically extract and analyse empirical CRS studies. We focus explicitly on how UX is conceptualised and operationalised across application domains, how adaptivity and LLM integration influence evaluation design, and what methods and instruments are employed. By comparing and categorising UX constructs and tools, we reveal which evaluation practices are prevalent, underutilised, or misaligned with the demands of generative CRS. Our synthesis bridges gaps between conceptual reviews and fragmented empirical work, offering actionable insights for CRS researchers and designers seeking to improve UX measurement and design practices.

\begin{table}[h]
\small
\centering
\caption{Comparison of 13 Prior Survey and Review Papers on Conversational and LLM-Powered Recommender Systems, examining their treatment of UX, empirical synthesis, evaluation methods, and LLM-specific focus. This table highlights key gaps in prior reviews that motivate the present systematic analysis.}
\label{tab:survey_comparison}
\begin{tabular}{p{2.8cm} p{2.8cm} p{1.8cm} p{1.8cm} p{1.8cm} p{1.8cm}}
\toprule
\textbf{Study} & \textbf{Primary Focus} & \textbf{UX Discussed} & \textbf{Empirical Synthesis} & \textbf{Eval. Methods Covered} & \textbf{LLM-Specific} \\
\midrule
Gao et al. (2021)~\cite{gao2021advances} & CRS Architectures & No & No & Yes & No \\
Lei et al. (2020)~\cite{lei2020conversational} & CRS Formulation \& Eval & No & No & Yes & No \\
Zaidi et al. (2024)~\cite{zaidi2024review} & CRS Techniques \& Applications & No & No & Yes & No \\
Pramod \& Bafna (2022)~\cite{pramod2022conversational} & CRS Tools \& Adoption & No & No & Yes & No \\
Jannach et al. (2023)~\cite{jannach2023evaluating} & Evaluation Framework & Yes & No & Yes & No \\
Wang et al. (2023)~\cite{wang2023rethinking} & Evaluation Critique & Yes & No & Yes & No \\
Li et al. (2023)~\cite{li2023conversation} & Holistic CRS Design & No & No & Yes & No \\
Lian et al. (2024)~\cite{lian2024recai} & LLM Toolkit (RecAI) & No & No & Yes & Yes \\
Wu et al. (2024)~\cite{wu2024survey} & LLM-RS Survey & Yes & No & Yes & Yes \\
Li et al. (2024)~\cite{li2024large} & LLM-RS Pipeline & No & No & Yes & Yes \\
Deldjoo et al. (2024)~\cite{deldjoo2404review} & LLM-RS Eval Taxonomy & Yes & No & Yes & Yes \\
Huang et al. (2025)~\cite{huang2025agentic} & Agentic RS Vision & No & No & No & Yes \\
Hou et al. (2024)~\cite{hou2024large} & LLMs as Rankers & No & No & Yes & Yes \\
\bottomrule
\end{tabular}
\end{table}

\section{Methodology}
\label{sec:methodology}

This review follows the PRISMA (Preferred Reporting Items for Systematic Reviews and Meta-Analyses) guidelines~\cite{moher_preferred_2009}, aiming to enhance reproducibility, transparency, and methodological rigour in synthesising UX evaluations of CRS. Our goal was to examine how CRS UX is conceptualised, operationalised, and measured across different application domains, adaptivity types, and system architectures, including LLM-based implementations.

\subsection{Review Protocol and Inclusion Criteria}
\label{sec:review-protocol}

We applied inclusive criteria to capture the diversity of CRS research. Studies were included if they: (1) implemented a CRS involving natural language interaction, (2) reported empirical UX findings using self-report, behavioural, or third-party methods, and (3) focused on any application domain or evaluation setting (lab, field, simulated). Exclusion criteria were: (1) studies lacking user evaluation (e.g., offline experiments only), (2) conceptual or design-only papers without empirical testing, and (3) technical implementations without user interaction or feedback. Only peer-reviewed conference and journal articles written in English were considered.

\subsection{Literature Search and Screening}
\label{sec:search-screening}

We conducted a comprehensive literature search across six major academic databases: ACM Digital Library, IEEE Xplore, ScienceDirect, SpringerLink, Scopus, and Web of Science. These sources were selected for their broad coverage of human-computer interaction (HCI), recommender systems, and conversational interfaces. The initial search was conducted in September 2023 and later updated in May 2025. To capture a wide spectrum of relevant research, we used the following search string:

\begin{lstlisting}
(convers* OR dialog* OR "speech" OR "voice assistant" OR "voice-enabled" OR 
"voice agent" OR "voice-based" OR "voice-activated" OR "spoken-language" OR 
chatbot OR chatterbot OR "large language model" OR LLM) 
AND ("User Experience" OR UX) AND recommend*
\end{lstlisting}

No date restrictions were applied, and Boolean operators were used to maximise recall. After deduplication, a two-phase screening procedure was employed, following PRISMA guidelines~\cite{moher_preferred_2009}. In Phase 1, titles and abstracts were screened for relevance against predefined inclusion criteria. In Phase 2, full texts were retrieved and assessed independently by multiple reviewers.

Screening was conducted by three independent reviewers. For the initial screening phase (records published up to September 2023), pairwise inter-rater agreement was measured using Cohen’s $\kappa$, with values ranging from 0.625 to 0.85 across title–abstract and full-text stages, indicating substantial to almost perfect agreement~\cite{mchugh2012interrater}. During the updated screening (October 2023 to May 2025), $\kappa = 0.66$ for the title–abstract phase and $\kappa = 0.88$ for the full-text phase were achieved. Disagreements were resolved through consensus discussions. Rayyan~\cite{ouzzani2016rayyan}, a dedicated web-based screening tool, was used to manage blinded decisions and track reviewer disagreements systematically.

\subsection{Data Extraction and Coding}
\label{sec:data-extraction}

A structured data extraction sheet was created and iteratively refined. Each included article was coded for the following fields: author(s), publication year, CRS domain (e.g., movie, music, nutrition), evaluation context (lab, field, remote), presence of adaptivity (adaptive vs.\ nonadaptive), use of LLMs or generative models, UX constructs evaluated, measurement tools used, and participant sample size.

UX constructs were extracted using the authors’ terminology, then categorised into broader UX types: 
\begin{itemize}
    \item \textbf{Affective:} enjoyment, surprise, trust, empathy
    \item \textbf{Cognitive:} usefulness, clarity, novelty, informativeness
    \item \textbf{Relational:} engagement, agency, control
    \item \textbf{Task-oriented:} satisfaction, efficiency, recommendation quality
\end{itemize}

Instruments and evaluation methods were coded into:
\begin{itemize}
    \item \textbf{SM (Self-report Measures):} Likert scales, interviews, NPS, UEQ
    \item \textbf{BM (Behavioural Measures):} dialogue length, error rate, task completion
    \item \textbf{EM (External Measures):} third-party annotations
\end{itemize}

Each study’s coding was verified by at least two researchers. Discrepancies were reconciled through iterative checking.

\subsection{Analysis and Synthesis Strategy}
\label{sec:synthesis-strategy}

We adopted a structured narrative synthesis aligned with the four research questions. For RQ1, we summarised and compared the UX constructs reported across studies, grouped by application domain. RQ2 was addressed by mapping evaluation tools and study designs to reveal dominant practices and underutilised methods. For RQ3, we categorised CRS adaptivity features and examined how these were evaluated in relation to UX outcomes. Finally, RQ4 was addressed through targeted analysis of LLM-powered CRS papers, triangulated with recent survey literature to surface unique UX challenges.
Due to the heterogeneity of UX constructs, evaluation tools, and study designs, we did not conduct quantitative meta-analysis. Instead, we focused on conceptual synthesis, identifying methodological trends, gaps, and design tensions. Descriptive summaries and tabular mappings supported comparison across studies.

\section{Results: Reviewed CRS UX Studies}
\label{sec:results}
This section presents findings from the included empirical studies evaluating user experience in CRSs. We have structured this section sequentially around our four research questions. We begin with an overview of all the articles.

\subsection{Article Overview}
\begin{figure}[tbph]
\centering
\begin{adjustbox}{max width=0.85\linewidth}
\begin{tikzpicture}[
    node distance=1.0cm and 1.0cm,
    mainnode/.style={
        rectangle, draw=black, thick, line width=1pt, 
        align=center, text width=3.2cm, rounded corners, 
        minimum height=1cm, fill=white, 
        font=\sffamily\normalsize\bfseries
    },
    label/.style={
        ellipse, draw=black, thick, line width=1pt, 
        align=center, fill=gray!20,
        font=\sffamily\normalsize\bfseries,
        minimum height=1.3cm, text width=2cm
    },
    every path/.style={
        ->, >=Stealth, line width=1pt
    }
]

\node[label] at (-4, 0) {Searching};
\node[label] at (-4, -6) {Screening};
\node[label] at (-4, -9.5) {Eligibility};
\node[label] at (-4, -13) {Included};

\node[mainnode] (acm) {ACM\\(n = 41)};
\node[mainnode, right=of acm] (ieee) {IEEE\\(n = 48)};
\node[mainnode, right=of ieee, xshift=-0.3cm] (sd) {ScienceDirect\\(n = 17)};
\node[mainnode, right=of sd] (springer) {Springer\\(n = 71)};
\node[mainnode, right=of springer] (scopus) {Scopus\\(n = 199)};
\node[mainnode, right=of scopus] (wos) {WOS\\(n = 84)};

\node[mainnode, below=2.5cm of sd] (id) {Records Identified\\(n = 460)};
\draw (acm) |- (id);
\draw (ieee) |- (id);
\draw ([yshift=-0.05cm]sd.south) -- ([yshift=0.05cm]id.north);
\draw (springer) |- (id);
\draw (scopus) |- (id);
\draw (wos) |- (id);

\node[mainnode, below=1.3cm of id] (dup) {Records after deduplication\\(n = 356)};
\node[mainnode, below=1.3cm of dup] (screened) {Title--Abstract Screened\\(n = 356)};
\node[draw=none, right=of screened, align=left, font=\sffamily\small] (excluded1) {Excluded\\(n = 306)};
\node[mainnode, below=1.3cm of screened] (fulltext) {Full-Text Screened\\(n = 50)};
\node[draw=none, right=of fulltext, align=left, font=\sffamily\small] (excluded2) {Excluded with Reasons\\(n = 27)};
\node[mainnode, below=1.3cm of fulltext] (included) {Articles Included\\in Review\\(n = 23)};

\draw (id) -- (dup);
\draw (dup) -- (screened);
\draw (screened) -- (fulltext);
\draw (fulltext) -- (included);
\draw (screened) -- (excluded1);
\draw (fulltext) -- (excluded2);

\end{tikzpicture}
\end{adjustbox}
\caption{PRISMA flow diagram illustrating the systematic screening, eligibility, and inclusion of studies in this review.}
\label{fig:prisma}
\end{figure}
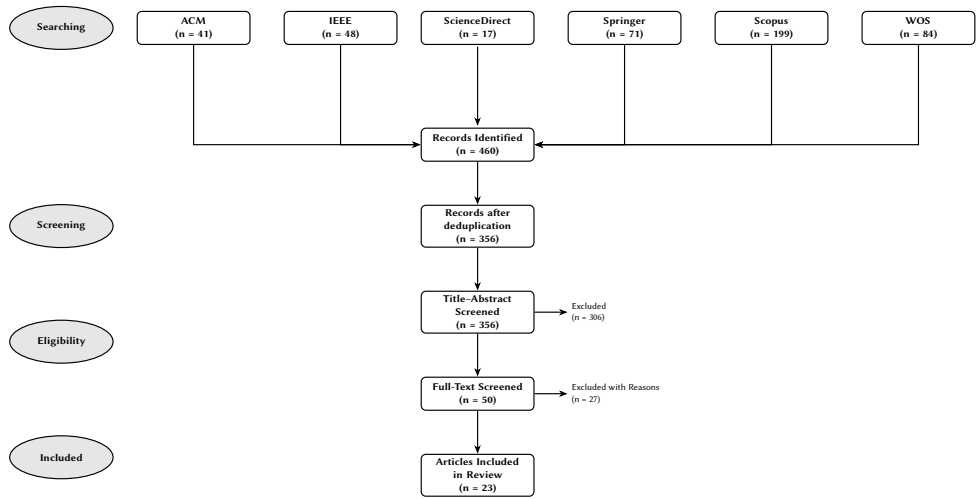

The PRISMA in Figure~\ref{fig:prisma} presents the flow of records from initial identification to final inclusion. After removing duplicates, 356 unique records were screened at the title and abstract level. Of these, 50 articles were selected for full-text review. Following exclusion of 27 papers that did not meet the inclusion criteria (e.g., insufficient UX data, non-conversational systems), 23 articles were included in our final synthesis. These studies span diverse domains, including e-commerce, music, movies, restaurants, nutrition, and domain-independent applications. Table~\ref{tab:summary_table} provides a detailed summary of each article, including domain, UX metrics, evaluation methods, and key findings.

\subsection{UX Dimensions and Domain Variation (RQ1)}
We examined what UX dimensions were evaluated across the studies and how these dimensions varied by application domain. A total of 71 distinct UX metrics were identified, spanning cognitive, affective, and behavioural dimensions. These metrics were extracted from both self-reported and behavioural measures, with a minority using human annotators who evaluated specific dialogues that occurred between a CRS and other users. We refer to those methods as external measures.

\paragraph{\textbf{Most Frequently Assessed Metrics.}}
User satisfaction was the most commonly reported UX metric, appearing in 18 of 23 studies. Other frequently evaluated dimensions included personalisation (9 studies), usefulness (7), trust (6), ease of use (6), and recommendation quality (6). Affective and reflective constructs such as curiosity, self-reflection, emotional support, and agency were reported in a smaller number of studies, predominantly those involving LLM-powered systems~\cite{yun2025user,gutierrez2023videolandgpt}.

\paragraph{\textbf{Domain-Specific Trends.}}
UX priorities varied systematically across application domains. E-commerce studies (n = 7) often focused on transactional UX factors including purchase intention, initiative, trust, and privacy. Music and movie CRS (n = 10) emphasised hedonic and affective UX such as enjoyment, novelty, serendipity, and transparency. Restaurant-focused studies (n = 3) featured constructs related to interaction quality, perspicuity, and usability. Nutrition systems (n = 2) assessed longer-term UX goals including loyalty, accessibility, and usage frequency. Domain-independent systems (n = 1) prioritised performance-oriented constructs such as interaction cost, recommendation accuracy, and efficiency. We next examined how these UX dimensions were measured and what tools and methodologies were employed to evaluate them across the included studies.

\small        
  \setlength{\tabcolsep}{4pt}  

  \begin{longtable}{
      p{0.05\textwidth}   
      p{0.08\textwidth}   
      p{0.20\textwidth}   
      p{0.25\textwidth}   
      p{0.28\textwidth}   
    }
    \caption{Summary of Article Characteristics: UX Metrics, Evaluation Methods, and Key Findings. This table presents data extracted from 23 empirical studies included in the review. Each row summarises the application domain, the UX metrics evaluated, the empirical methods and tools used (e.g., behavioural measures, surveys, interviews), and key UX-related findings. Evaluation types are denoted as SM (self-report measures), BM (behavioural measures), and EM (external measures, i.e., UX appraised by non-user raters). This synthesis supports the comparative analysis of UX evaluation practices across classical and LLM-powered CRS.}

    \label{tab:summary_table} \\

    \toprule
    \textbf{Study}
    & \textbf{Domain}
    & \textbf{UX Metrics}
    & \textbf{Evaluation Methods \& Tools}
    & \textbf{Key Findings} \\
    \midrule
    \endfirsthead

    \toprule
    \textbf{Study}
    & \textbf{Domain}
    & \textbf{UX Metrics}
    & \textbf{Evaluation Methods \& Tools}
    & \textbf{Key Findings} \\
    \midrule
    \endhead

    \midrule \multicolumn{5}{r}{\textit{Continued on next page}} \\
    \midrule
    \endfoot

    \bottomrule
    \endlastfoot

\cite{yun2025user}
& Music
& Agency; Emotional Support; Exploration; Personalisation; Reflection.
& \textbf{SM}: Three-week diary study (\(n = 12\)) involving free-form conversations with GPT-based CRS; daily diary entries and post-study interviews designed by researchers to elicit affective and reflective user experiences;  
Analysis: inductive thematic analysis \cite{braun_using_2006}; narrative synthesis of diary and interview content.
& CRS fostered exploration and self-reflection, providing emotional support during everyday life. However, users also experienced frustration due to limited memory and generic responses. Perceived personalisation was linked to anthropomorphic behaviours and follow-up questions. No standardised UX scales or statistical analyses were used. \\

\cite{ma2024effect}
& E-commerce
& Supportive; Easy; Efficient; Clear; Exciting; Interesting; Inventive; Leading Edge.
& \textbf{SM}: Between-subjects online study (\(n = 184)\) comparing proactive vs. passive interaction schemes; UX evaluated using the UEQ-S \cite{schrepp_design_2017}; proactive group received rule-based prompts for decision aids (e.g., compare, critique, more details); pre- and post-interaction questionnaires measured decision-making style and meta-intents.
& No significant differences in UEQ-S scores across schemes, though proactive group reported higher means on 'interesting' and 'exciting' (d = .159, .120), while passive scored better on 'clear' and 'easy' (d = –.155, –.131); prompts significantly increased usage of decision aids (e.g., “check details”: d = 1.45, p < .001; “compare in dialog”: d = .70, p < .001). \\

\cite{kraus2024pilot}
& Restaurant
& Efficiency; Perspicuity; Attractiveness; Novelty; Stimulation; Dependability; Stimulation.
& \textbf{BM}: dialogue logs (conversation duration; moderation actions; utterance counts; group task performance);
  \textbf{SM}: post‐interaction surveys  (\(n = 21)\) using UEQ (7‐point Likert, long form) and BFI‐10 personality scale. 
  Analysis: Wilcoxon tests; Bonferroni‐corrected Spearman’s correlations; descriptive statistics; qualitative content coding (Cohen’s \(\kappa = 0.683\)).
& No significant UEQ differences between leader vs. follower conditions (\(p > 0.05\)). Conversation‐leading ↓ interaction time (728s vs. 964s, \(p = 0.09\)); ↑ dominance gap (8.20 vs. 2.87, \(p = 0.06\)); ↓ moderation acts (\(p = 0.04\)). Follower strategy ↑ task success (35.57 vs. 32.71 points). Openness positively correlated with perspicuity (\(\rho = 0.65,\;p = 0.01\)) and dependability (\(\rho = 0.63,\;p = 0.02\)) in leader condition. \\

\cite{thom2024nutria}
& Nutrition
& Satisfaction; Effort; Loyalty; Usefulness; Accuracy; Accessibility; Frequency of Use.
& \textbf{SM}: Online survey administered after demo and independent use (\(n = 50)\); custom-built questionnaires based on CSAT, CES, NPS, and usage metrics; response options used 5-point Likert scale and frequency scale; 
Analysis: formula-based score computation from raw survey data (no inferential statistics).
& CSAT: 96\% (Mean = 4.60); CES: 90\% (Mean = 4.42); NPS: 92\% (Mean = 4.38); Accuracy: 94\% (Mean = 4.56); Use frequency: 5 times/week; Satisfaction with recommendations: 98\%. \\

\cite{gutierrez2023videolandgpt}
& Movie
& Enjoyment; Personalisation; Relevance; Satisfaction.
& \textbf{SM}: Between-subjects study (\(n = 27)\) comparing two CRS versions (personalised vs. non-personalised) across five interaction tasks;  
  post-task questionnaire included Likert-scale items (5-point) measuring user satisfaction and perceived experience;  
  ranking-based accuracy (nDCG@9) and hit-rate (HR@9) computed for each version and task;  
  Analysis: descriptive statistics; Pearson’s \(r = 0.26\) between nDCG@9 and user preferences.
& Personalised system showed higher average nDCG@9 (0.4273 vs. 0.3880) and HR@9 (0.78 vs. 0.74);  
  40\% of personalised users agreed with all positive Likert statements vs. 10\% in non-personalised;  
  ~22\% of recommendations were not on the platform, raising fairness concerns. \\

    \cite{ma2023initiative}
    & E-commerce
    & Information sufficiency; Initiative; Interaction adequacy; Efficiency; Interest; Accuracy; Satisfaction; Usefulness.
    & \textbf{BM}: count/duration of interactions (before vs. after initiative transfer);  
      \textbf{SM}: pre/post 5‐point Likert surveys based on Decision‐making Style, Sense of Agency, Initiative Preference, UEQ‐S, ResQue;  
      Analysis: descriptives; independent‐samples \(t\)-tests; chi‐square; two‐way MANOVA.
    & Switching to user‐initiated mode produced fewer interactions (\(t(86.20) = -2.97,\;p=0.012\)) and shorter duration (\(t(141) = -1.36,\;p=0.175\)). Initiative transfers decreased excitement (\(F(1,143)=7.07,\;p=0.009\)) and interest (\(F(1,143)=5.25,\;p=0.023\)). Chi‐square for transfer occurrence: \(\chi^2(1,N=143)=24.40,\;p<0.001\). \\

    \cite{el2023sentiment}
    & E-commerce
    & Personalisation; Sentiment Analysis; User Satisfaction.
    & \textbf{BM}: sentiment classifiers evaluated by F1‐score;  
      \textbf{SM}: post‐interaction survey (0–10 scale) adapted from Teo et al., plus custom questions for personalisation;  
      Analysis: compute F1 metrics; A/B tests comparing satisfaction (with SA vs. without SA).
    & Incorporating sentiment analysis raised satisfaction: SA group 9.13/10 vs. non‐SA 8.41/10. F1‐scores: BERT (0.76→0.85); GPT (0.72→0.78); ELMO (0.73→0.82). \\

    \cite{rana2024user}
    & Restaurant
    & Diversity; Ease of Use; Effectiveness; Future Intention; Personalisation; Relevance; Satisfaction; Serendipity; Unexpectedness.
    & \textbf{BM}: interaction logs (failures, effort, flow, recommendation quality);  
      \textbf{SM}: post‐interaction survey (5‐point Likert); custom questionnaires;  
      Analysis: descriptives; one‐way \(t\)-test; one‐sided \(z\)-test.
    & Failure rate: NP‐Rec 56.1\% vs. P‐Rec 50\% (\(p=0.039\)). Avg. interactions: NP‐Rec 3.5 vs. P‐Rec 2.8 (\(p=0.003\)). Relevance: P‐Rec 85\% vs. NP‐Rec 78\%. Future interest: P‐Rec 68\% vs. NP‐Rec 59\%. Kappa (ease of use): P‐Rec 0.918; NP‐Rec 0.857. \\

    \cite{siro2023understanding}
    & Movie
    & Interest arousal; Recommendation quality (precision, recall, F1); Efficiency; User Satisfaction.
    & \textbf{EM}: Wizard‐of‐Oz dialogue annotations by trained personnel; crowdsourced MTurk evaluations (3‐point, 4‐point, 5‐point Likert);  
      Analysis: marginal distributions; box plots; correlations; regression; classification models.
    & Turn‐level relevance correlated with satisfaction (\(\rho=0.610\); Spearman). Dialogue‐level task completion (\(\rho=0.599\)) and interest (\(\rho=0.622\)) predicted satisfaction. Turn‐level prediction: Pearson \(r=0.734\). Dialogue‐level \(r=0.796\). Random forest RMSE = 0.768, \(\rho=0.734\). \\

    \cite{jin_birds_2022}
    & E-commerce
    & Future Intentions; Perceived Friendliness; Product Attitudes; User Satisfaction.
    & \textbf{SM}: post‐interaction surveys (5‐point Likert) for satisfaction, perceived friendliness, product attitudes, chatbot intention;  
      Analysis: PROCESS macro Models 1 \& 7; Johnson–Neyman technique.
    & Extrovert‐matched chatbots increased satisfaction and friendliness among extroverts (\(F(1,286)=52.52,\,p<0.001\)). Mediation: satisfaction → future intentions \& attitudes (Index = 0.35, SE = 0.13, 95 
    \\

    \cite{chung2022consumer}
    & E-commerce
    & Purchase Intentions; Psychological Distance; Usability.
    & \textbf{SM}: post‐interaction surveys using screenshots; scales for anthropomorphism, conversational relevance (custom), psychological distance, usability, purchase intention;  
      Analysis: two‐way ANOVA.
    & High relevance improved task proficiency (\(F(1,58)=56.58,\,p<0.01\)) and perceived intelligence (\(F(1,58)=24.30,\,p<0.01\)). High anthropomorphism increased perceived responsibility (\(F(1,58)=5.65,\,p=0.02\)). Purchase intentions: high relevance (\(F(1,58)=28.74,\,p<0.01\)). \\

    \cite{martina_narrative_2022}
    & Movie
    & Ease of Use; Interaction Adequacy; Intention to Use; Recommendation Accuracy; User Control; User Satisfaction.
    & \textbf{BM}: interaction logs— interaction duration, HitRate@K, preference count; \textbf{SM}: post‐interaction survey (5‐point Likert) based on ResQue;  
    Analysis: HitRate@1/@3; descriptive comparisons.
    & HitRate@1: Objective 78 \% vs. Objective + Subjective 88.5\%. HitRate@3: Objective 96\% vs. Objective + Subjective 98.1\%. Ease of Use: Objective 4.24 vs. Objective + Subjective 4.26. User Control: Objective 4.28 vs. Objective + Subjective 4.19. Interaction Adequacy: Objective 4.28 vs. Objective + Subjective 4.15. Recommendation Accuracy: Objective 4.30 vs. Objective + Subjective 4.50. User Satisfaction: Objective 4.20 vs. Objective + Subjective 4.40. \\

    \cite{silva_polite_2022}
    & E-commerce
    & Linguistic Politeness.
    & \textbf{BM}: automated metrics—Politeness Score, BLEU, ROUGE, METEOR;  
      \textbf{EM}: dialogue annotations (1–3 scale; binary politeness/grammar);  
      Analysis: compute automated metrics; comparative statistics.
    & Rewrite methods outperformed generative ones in politeness and content retention. Politeness scores: RW‐Enron 82.24; RW‐Fashion 79.76; RW‐Fashion‐C 78.42; RW‐Mixed 80.70. RW‐Fashion‐C \& RW‐Mixed had significantly better BLEU/ROUGE/METEOR. \\

    \cite{cai_enhancing_2022}
    & Music
    & Confidence; Control; Discovery; Diversity; Ease of Use; Interaction Adequacy; Interaction Efficiency; Novelty; Satisfaction; Serendipity; Transparency; Trust.
    & \textbf{BM}: \# interactions; dialogue turns; session duration; \# critiques; accepted recommendations;  
      \textbf{SM}: post‐task survey (7‐point Likert) based on ResQue \& Matt et al.;  
      Analysis: SEM; Kruskal–Wallis; Mann–Whitney U; Mixed ANOVA; post‐hoc.
    & Exploration‐oriented tasks led to higher interaction (\(p<0.001\)) and reduced satisfaction (\(p<0.01\)) vs. basic tasks. Cascading critiquing ↑ diversity (\(p<0.05\)); progressive critique ↑ serendipity (\(p<0.01\)). Significant differences across critiquing techniques for serendipity, diversity, adequacy (\(p<0.05\)). \\

    \cite{samagaio_chatbot_2021}
    & Nutrition
    & Personalisation; Sentiment; Usability; User Satisfaction.
    & \textbf{BM}: precision/recall/F1 via DIET; sentiment analysis accuracy via VADER; \textbf{SM}: post‐interaction surveys: SUS, UEQ, CUQ, Single Ease Question;  
      Analysis: descriptive statistics; \(t\)‐tests; one‐way ANOVA.
    & SUS = 84.7 ± 8.4; CUQ = 82.9 ± 8.6. Sentiment accuracy: Spearman’s \(\rho=0.67\), Pearson’s \(r=0.68\), MAE = 0.85. User satisfaction ↑ significantly with personalised recommendations (\(F(1,58)=27.45,\,p<0.001\)). \\

    \cite{iovine_investigation_2021}
    & Domain-independent
    & Interaction Cost; Recommendation Quality; User Satisfaction.
    & \textbf{BM}: interaction logs— \#questions; interaction time; query density; recommendation accuracy; MAP; \textbf{SM}: post‐interaction surveys: Likert; Analysis: MANOVA.
    & Natural‐language interface resulted in fewer questions but longer interaction time vs. button interface. Mixed mode ↑ efficiency \& recommendation quality (\(p<0.05\)), with highest accuracy = 0.63 and MAP = 0.55. \\

    \cite{anastasia_designing_2021}
    & E-commerce
    & Privacy Risk; Social Presence; Trust; Usability.
    & \textbf{BM}: time per task; completion rate;  
      \textbf{SM}: post‐interaction surveys \& interviews (custom);  
      Analysis: descriptive statistics.
    & High privacy (6.2/7) and social presence (6.7/7) scores; usability = 94.5/100; 100 
    \\

    \cite{jin_key_2021}
    & Music
    & Confidence; Engagement; Humanness; Interaction Adequacy; Novelty; Perceived Ease of Use; Perceived Usefulness; Transparency; Trust; User Control; User Satisfaction.
    & \textbf{BM}: \# interactions; duration; recommendation acceptance rate; task success rate; 
      \textbf{SM}: post‐interaction surveys (7‐point Likert) based on ResQue model \& custom items;  
      Analysis: SEM.
    & SEM: Perceived usefulness (\(\beta=0.52,\,p<0.01\)), ease of use (\(\beta=0.45,\,p<0.01\)), transparency (\(\beta=0.43,\,p<0.01\)) → user satisfaction. Novelty (\(\beta=0.35,\,p<0.05\)) \& interaction adequacy (\(\beta=0.32,\,p<0.05\)) → user control. Recommendation accuracy = 90 
    \\

    \cite{fernando_enhancing_2021}
    & Music
    & Effectiveness; Efficiency; Persuasiveness; Transparency; Trustworthiness; User Satisfaction.
    & \textbf{SM}: pre/post interaction surveys using standardized questionnaires based on Tintarev;  
      Analysis: descriptive statistics.
    & Explanation feature ↑ user satisfaction, perceived transparency, trustworthiness, effectiveness, persuasiveness. Recommendation accuracy = 90.48 
    \\

    \cite{kraus_comparison_2020}
    & Restaurant
    & Cognitive Demand; Habitability; Likeability; Motivation to Interact; Response Accuracy; User Satisfaction.
    & \textbf{SM}: post‐interaction surveys (7‐point Likert) using User Acceptance Scale, SASSI, Motivation Scale;  
    Analysis: one‐way ANOVA; post‐hoc \(t\)‐tests.
    & Both proactive strategies (explicit \& implicit) ↑ satisfaction vs. reactive (\(F(2,16)=3.65,\,p<0.05\)). Implicit strategy ↑ accuracy \& habitability (\(F(2,16)=4.81,\,p<0.05\)). \\

    \cite{iovine_humanoid_2020}
    & Movie
    & Confidence; Ease of Use; Interaction Adequacy; Overall Satisfaction; Recommendation Accuracy; Transparency; Trust; Use Intentions; Usefulness; User Control.
    & \textbf{BM}: interaction cost; \# questions; query density; recommendation accuracy; MAP;  
      \textbf{SM}: post‐interaction surveys (5‐point Likert) based on ResQue model.
      Analysis: Wilcoxon signed‐rank.
    & No significant difference in overall satisfaction. Recommendation accuracy: Smartphone 3.80 vs. Robot 3.65 (MAP: no difference). Robot ↑ enjoyment (\(p=0.046\)) but ↑ irritation (\(p=0.043\)). Trust: Robot 3.45 vs. Smartphone 3.75 (\(p<0.05\)). Transparency improved with Robot (\(p<0.05\)). \\

    \cite{pecune_model_2019}
    & Movie
    & Decision Confidence; Intention to Return; Intention to Watch; Perceived Effort; Perceived Usefulness; Recommendation Quality; Transparency; User Control; User Satisfaction.
    & \textbf{SM}: post‐interaction surveys (5‐point Likert) based on prior works; Analysis: MANOVA; two‐way ANOVA.
    & Social explanations ↑ recommendation quality (\(F=12.45,\,p<0.01\)) \& personalised ratings (\(F=15.32,\,p<0.01\)). Decision confidence ↑ (\(F=10.87,\,p<0.01\)); overall satisfaction ↑ (\(p<0.01\)). Transparency, perceived effort, and user control ↑ (\(p<0.05\)). Return intention ↑ (\(p<0.05\)). \\

    \cite{lee2017enhancing}
    & Movie
    & Intimacy; Intention to Use; Interactional Enjoyment; Trust; User Satisfaction.
    & \textbf{SM}: post‐interaction surveys (7‐point Likert) based on CASA \& Uncertainty Reduction Theory;  
      Analysis: two‐way ANOVA; PLS regression.
    & Self‐disclosure (\(F=10.99,\,p<0.01\)) \& reciprocity (\(F=17.47,\,p<0.001\)) ↑ satisfaction. Perceived trust \& enjoyment mediated effects (\(p<0.05\)). Reciprocity was a stronger predictor of relationship building than self‐disclosure. \\
  \end{longtable}
  \vspace*{-0.8\baselineskip}

\subsection{Evaluation Methods and Tools Used in CRS UX Studies (RQ2)}
UX evaluation in CRS research is predominantly conducted using self-reported measures, found in 21 of the 23 studies. These methods typically involved post-interaction survey questionnaires using Likert-type scales (5-point or 7-point), capturing constructs such as satisfaction, usability, trust, novelty, and intention to use. Many studies adapted validated instruments such as UEQ (User Experience Questionnaire), UEQ-S, ResQue, SUS, or domain-specific scales (e.g., CSAT, CES, NPS).
Behavioural measures were reported in 12 studies. These included logging the number of dialogue turns, interaction durations, recommendation acceptance rates, moderation actions, and task completion. For example, Ma et al. \cite{ma2023initiative} compared interaction lengths across initiative strategies; Gutierrez et al. \cite{gutierrez2023videolandgpt} measured nDCG@9 and HitRate@9 for CRS versions with and without personalisation.
External measures were rarely used. Only two studies adopted third-party annotations or crowd-worker ratings. Siro et al. \cite{siro2023understanding} employed Wizard-of-Oz simulations and MTurk evaluations to link dialogue turn quality to satisfaction. Silva et al. \cite{silva_polite_2022} annotated politeness and grammar scores externally to benchmark generative responses.
Qualitative methods were used in three studies. These included interviews, open-ended reflections, and thematic coding of diary entries. For instance, Yun and Lim \cite{yun2025user} conducted a three-week diary study on GPT-based CRS, applying Braun and Clarke’s (2006) inductive thematic analysis. Kraus et al. \cite{kraus2024pilot} combined content coding with personality correlations to understand group dynamics in multi-party CRS.
Regarding tool adoption, 13 studies explicitly cited validated frameworks or tools. The remaining studies either designed novel items or minimally described their instrument sources, reducing replicability.
Inferential analyses included t-tests (7 studies), ANOVAs (6), MANOVAs, Wilcoxon, and Spearman correlations. A smaller subset used structural models (e.g., SEM in 2 studies), classification models, or effect size reporting (e.g., Cohen’s d, Pearson’s r). However, several recent works, particularly those using generative CRS, reported descriptive insights without statistical testing, limiting the generalisability of their findings.
Taken together, the findings suggest that while survey-based self-reporting dominates CRS UX evaluation, a growing number of studies are augmenting this with behavioural logging and qualitative narratives. Having examined how UX was measured, we next investigated how adaptivity and personalisation influenced CRS design and evaluation.

\subsection{Adaptive vs. Nonadaptive UX (RQ3)}
\label{sec:results-adaptivity}

We examined how CRS studies incorporated adaptivity and personalisation into their system design and UX evaluation strategies. We classified each of the 23 studies based on whether the system modified its behaviour dynamically during interaction. We considered a system adaptive if it adjusted its dialogue flow, initiative, or content generation in response to user input, preferences, or traits. Systems that included personalisation without runtime behavioural change were classified as nonadaptive.
We identified seven studies as fully adaptive~\cite{yun2025user, ma2024effect, kraus2024pilot, ma2023initiative, martina_narrative_2022, kraus_comparison_2020, pecune_model_2019} and one study as partially adaptive~\cite{gutierrez2023videolandgpt}. The remaining fifteen studies used fixed interaction logic and did not incorporate behavioural adaptation during runtime~\cite{rana2024user, silva_polite_2022, jin_birds_2022, cai_enhancing_2022, siro2023understanding, fernando_enhancing_2021, iovine_investigation_2021, iovine_humanoid_2020, lee_enhancing_2017, anastasia_designing_2021, samagaio_chatbot_2021, jin_key_2021, el-ansari_sentiment_2023, thom2024nutria, chung_consumer_2022}.

\paragraph{\textbf{Adaptivity Strategies.}}
Several systems supported initiative shifts based on user input or interaction style, including explicit toggling between proactive and reactive dialogue flows~\cite{ma2024effect, ma2023initiative, kraus_comparison_2020}. In group decision-making contexts, adaptive logic guided the assignment or alternation of agent roles, such as leader or supporter, based on conversational dynamics and user preference inputs~\cite{kraus2024pilot}. Other systems used natural language inputs to infer user goals and adjust recommendations dynamically, including narrative-driven content adaptation~\cite{martina_narrative_2022} and affect-sensitive interaction design using goal-oriented versus open-ended prompts~\cite{yun2025user}. Pecune et al.~\cite{pecune_model_2019} implemented real-time personalisation of social explanations by aligning system messages with users' Big Five personality traits. These systems employed a range of adaptivity techniques, including policy-based switching, cue-triggered behaviour modulation, and language model-driven personalisation routines. Researchers often examined adaptive control through experimental manipulations of initiative style, cue types, or framing conditions.

\paragraph{\textbf{Personalisation Depth.}}
Nineteen studies incorporated personalisation, although its function and depth varied widely. Many nonadaptive systems used static personalisation methods, such as filtering content or priming response tone based on user traits~\cite{jin_birds_2022, lee_enhancing_2017}, sentiment~\cite{el-ansari_sentiment_2023}, or domain-specific constraints like dietary needs~\cite{thom2024nutria}. These systems typically personalised surface-level content without modifying the interaction logic. In contrast, adaptive systems integrated user traits, preferences, or affective signals directly into dialogue planning, modulating recommendation content, explanation strategies, or initiative dynamics~\cite{ma2024effect, yun2025user, pecune_model_2019}. While several studies acknowledged the distinction between preference-based and behavioural personalisation, few articulated this difference explicitly. Only a minority of studies evaluated the UX impact of personalisation mechanisms, with most relying on between-condition comparisons or post-interaction feedback to assess user alignment.

\paragraph{\textbf{Evaluation Coverage.}}
Evaluation approaches also differed by adaptivity. Studies featuring adaptive CRS more frequently assessed interactional UX metrics such as perceived control, initiative alignment, and engagement, often through experimental manipulations or mixed methods~\cite{kraus_comparison_2020, ma2024effect, martina_narrative_2022}. Nonadaptive systems tended to rely on post hoc ratings of satisfaction, clarity, or content relevance, often collected using static Likert scales or sentiment annotations~\cite{rana2024user, silva_polite_2022, siro2023understanding, fernando_enhancing_2021}. Only a small number of studies explicitly evaluated the adaptivity mechanism itself, typically through comparative designs testing alternate system behaviours.
We found that adaptive systems supported more complex evaluation logic and broader UX metric coverage, while nonadaptive systems focused primarily on interface impressions and content quality. Although most studies incorporated some form of personalisation, only a subset used it to influence system behaviour dynamically during interaction. We now turn to the subset of studies that utilised LLMs, revealing emerging implementation trends and critical methodological limitations.

\begin{small}
\begin{longtable}{p{0.02\textwidth} p{0.21\textwidth} p{0.07\textwidth} p{0.09\textwidth} p{0.48\textwidth}}
\caption{Adaptivity and Personalisation in Reviewed CRS Studies. Studies are ordered by publication year. Adaptivity refers to dynamic system behaviour that responds to user input, feedback, or traits during the interaction. Personalisation refers to any tailoring of content, recommendations, or dialogue style based on user-specific information, including preferences, personality, or interaction history.} \label{tab:adaptivity_personalisation_crs} \\
\toprule
\textbf{\#} & \textbf{Study} & \textbf{Adapt.} & \textbf{Personal.} & \textbf{Evidence Summary} \\
\midrule
\endfirsthead

\multicolumn{5}{c}%
{\tablename\ \thetable\ -- \textit{Continued from previous page}} \\
\toprule
\textbf{\#} & \textbf{Study} & \textbf{Adapt.} & \textbf{Personal.} & \textbf{Evidence Summary} \\
\midrule
\endhead

\midrule \multicolumn{5}{r}{\textit{Continued on next page}} \\
\endfoot

\bottomrule
\endlastfoot

1 & Yun et al. (2025) \cite{yun2025user} & Yes & Yes & GPT-based music CRS supported affective adaptation using goal-directed vs. open-ended dialogue. \\
2 & Ma \& Ziegler (2024) \cite{ma2024effect} & Yes & Yes & Trait-sensitive proactive cues; dialogue initiative adjusted based on decision style. \\
3 & Kraus et al. (2024) \cite{kraus2024pilot} & Yes & Yes & CRS alternated between leader and follower roles in a multi-party restaurant recommendation task; individual preferences were modelled to facilitate group decision-making. \\
4 & Damian et al. (2024) \cite{thom2024nutria} & No & Yes & NutrIA mobile app used static logic but personalised nutrition based on user inputs. \\
5 & Rana et al. (2024) \cite{rana2024user}& No & Yes & Explores user preferences in critiquing CRS designs; no live adaptation. \\
6 & Ma \& Ziegler (2023) \cite{ma2023initiative} & Yes & Yes & CRS enabled user-controlled initiative switching with trait-based framing. \\
7 & El-Ansari \& Beni-Hssane (2023) \cite{el-ansari_sentiment_2023} & No & Yes & Sentiment-informed response tone in chatbot; no adaptive logic applied during interaction. \\
8 & Granada et al. (2023) \cite{gutierrez2023videolandgpt} & Partial & Yes & LLM-based movie CRS (VideolandGPT); prompts tailored to user traits, but lacks adaptive dialogue policy. \\
9 & Siro et al. (2023) \cite{siro2023understanding} & No & Yes & Predictive UX model trained on user features; system behaviour remained static. \\
10 & Jin \& Eastin (2022) \cite{jin_birds_2022} & No & Yes & Personality similarity influenced trust but did not modify system behaviour. \\
11 & Chung \& Han (2022) \cite{chung_consumer_2022} & No & Yes & Rule-based e-commerce chatbot with fixed scripted responses; personalised framing manipulated via warm vs. cold chatbot persona in experimental conditions. \\
12 & Martina et al. (2022) \cite{martina_narrative_2022} & Yes & Yes & Narrative CRS dynamically adapted recommendations from free-text preferences. \\
13 & Silva et al. (2022) \cite{silva_polite_2022} & No & No & Task-oriented politeness agent with scripted, non-tailored interactions. \\
14 & Cai et al. (2022) \cite{cai_enhancing_2022} & No & No & Task-oriented critiquing CRS; lacked adaptive updates or user models. \\
15 & Samagaio et al. (2021) \cite{samagaio_chatbot_2021} & No & No & Recipe CRS using rules and slots; no user-specific responses. \\
16 & Iovine et al. (2021) \cite{iovine_investigation_2021} & No & No & Natural language variants tested; interaction flow was fixed. \\
17 & Anastasia et al. (2021) \cite{anastasia_designing_2021} & No & Yes & Embodied e-commerce agent designed for social alignment; no adaptation to user behaviour. \\
18 & Jin et al. (2021) \cite{jin_key_2021} & No & Yes & CRS-Que used to analyse perceived quality dimensions; not used for adaptive control. \\
19 & Fernando et al. (2021) \cite{fernando_enhancing_2021} & No & Yes & Music recommendations explained via personality traits; system remained non-adaptive. \\
20 & Kraus et al. (2020) \cite{kraus_comparison_2020} & Yes & Yes & Dialogue initiative adapted to user input style; mixed-initiative strategy verified. \\
21 & Iovine et al. (2020) \cite{iovine_humanoid_2020} & No & Yes & Robot CRS with static preference input; no dynamic personalisation beyond domain. \\
22 & Pecune et al. (2019) \cite{pecune_model_2019} & Yes & Yes & Social explanations adapted in real-time to Big Five traits. \\
23 & Lee \& Choi (2017) \cite{lee_enhancing_2017} & No & Yes & Self-disclosure and reciprocity studied; CRS responses were non-adaptive. \\
\end{longtable}
\end{small}

\subsection{Evaluating LLM-Powered CRS (RQ4)}
\label{sec:results-rq4-llm}

Two studies in the review explicitly implemented CRS systems powered by LLMs: Yun et al.~\cite{yun2025user} and Granada et al.~\cite{gutierrez2023videolandgpt}. Both systems leveraged \textbf{OpenAI’s GPT} architecture for natural language generation, though their underlying recommendation logic and evaluation designs differed.
Yun et al.~\cite{yun2025user} introduced a GPT-based music CRS that supported affective adaptation through manipulation of dialogue modes (goal-directed vs.\ open-ended prompts). The system was evaluated using a between-subjects experimental design, with outcome variables including user trust, perceived friendliness, and intention to reuse. GPT-based generation was controlled via prompt framing; however, the system lacked structured explanation modules or user-facing transparency mechanisms.
Granada et al.~\cite{gutierrez2023videolandgpt} deployed a hybrid CRS architecture combining retrieval-based candidate generation with GPT-3.5-based free-form responses. Recommendations were presented via naturalistic dialogue generated from ranked items embedded within GPT prompts. The evaluation consisted of a comparative user study measuring perceived coherence, engagement, and dialogue naturalness. No formalised UX instrument was employed, and evaluation focused on qualitative user impressions.
Neither system incorporated turn-level UX instrumentation, explanation transparency controls, or adaptive modulation of verbosity. Both evaluations were limited to single-session interactions, with no longitudinal tracking of usability, satisfaction, or trust dynamics. Table~\ref{tab:implementation_summary} summarises architecture reporting and implementation tooling across all reviewed studies, illustrating the limited methodological coverage in current LLM-powered CRS evaluations.

\begin{table}
\small
\centering
\caption{Implementation Reporting and Tooling across Reviewed CRS Studies (2025--2017). This table summarises the extent of system architecture reporting and the implementation stack used in each study, with a focus on LLM involvement where relevant.}
\label{tab:implementation_summary}
\begin{tabular}{p{0.02\textwidth} p{0.23\textwidth} p{0.24\textwidth} p{0.43\textwidth}}
\toprule
\textbf{\#} & \textbf{Study} & \textbf{Architecture Reporting} & \textbf{Modelling/Implementation Stack} \\
\midrule
1 & Yun et al. (2025)~\cite{yun2025user} & Reported: GPT-based architecture & OpenAI GPT API, goal-directed vs. open-ended prompting, music domain KB integration \\
2 & Ma \& Ziegler (2024)~\cite{ma2024effect} & Reported: Wizard-of-Oz via Telegram & Predefined response dictionary, simulated Telegram interface using pyTelegramBotAPI + MySQL \\
3 & Kraus et al. (2024)~\cite{kraus2024pilot} & Reported: Wizard-of-Oz with slot-filling & Rule-based Telegram interface, pyTelegramBotAPI, MySQL cloud logging \\
4 & Damian et al. (2024)~\cite{thom2024nutria} & Partially reported & Static logic via nutrition DB, Android frontend, rule-based interactions \\
5 & Rana et al. (2024)~\cite{rana2024user} & Not reported & Study design focused on user critique responses; no CRS implementation described \\
6 & Ma \& Ziegler (2023)~\cite{ma2023initiative} & Reported: Modular, non-generative CRS & Google Dialogflow (intent classification), modular action-response with UI widget bindings \\
7 & El-Ansari \& Beni-Hssane (2023)~\cite{el2023sentiment} & Partially reported & Sentiment-informed response variation, rule-based outputs, JavaScript interface \\
8 & Granada et al. (2023)~\cite{gutierrez2023videolandgpt} & Reported: GPT-3.5 for response generation & GPT-3.5 used for NLG; candidate reranking fused into prompts \\
9 & Siro et al. (2023)~\cite{siro2023understanding} & Reported: Offline UX modelling only & XGBoost trained on post-interaction logs and features, no interactive system \\
10 & Jin \& Eastin (2022)~\cite{jin_birds_2022} & Partially reported & Scripted persona framing in Qualtrics; no real-time CRS logic described \\
11 & Chung \& Han (2022)~\cite{chung_consumer_2022} & Reported: Scripted persona variants & Warm/cold framing scripts for e-commerce scenario; no adaptive mechanisms \\
12 & Martina et al. (2022)~\cite{martina_narrative_2022} & Reported: Modular NLP pipeline & Dialogflow (intent recognition), CoreNLP (NER, sentiment), Doc2Vec (CB engine), Python backend \\
13 & Silva et al. (2022)~\cite{silva_polite_2022} & Reported: Scripted dialogue logic & Fixed politeness templates, no adaptive control \\
14 & Cai et al. (2022)~\cite{cai2022impacts} & Reported: Static form-based CRS & Slot-based critique flow, template logic, no runtime adaptation \\
15 & Samagaio et al. (2021)~\cite{samagaio_chatbot_2021} & Reported: Rule-based slot-filling & Fixed-response preference chatbot, no ML or NLP used \\
16 & Iovine et al. (2021)~\cite{iovine_investigation_2021} & Reported: Scripted dialogue variants & Dialogue variants used to test linguistic framing; no CRS engine described \\
17 & Anastasia et al. (2021)~\cite{anastasia_designing_2021} & Reported: Embodied scripted agent & Virtual avatar with scenario-driven, scripted interaction \\
18 & Jin et al. (2021)~\cite{jin2021key} & Not reported & CRS-Que questionnaire development only; no implemented agent \\
19 & Fernando et al. (2021)~\cite{fernando_enhancing_2021} & Reported: Template-based trait alignment & Static music explanations based on Big Five traits \\
20 & Kraus et al. (2020)~\cite{kraus_comparison_2020} & Reported: Alexa Skill implementation & FlaskAsk, Alexa SDK, proactive cue variants in controlled dialogue \\
21 & Iovine et al. (2020)~\cite{iovine_humanoid_2020} & Reported: Static NAO robot dialogue & Pre-programmed utterances on humanoid robot platform, MySQL logging \\
22 & Pecune et al. (2019)~\cite{pecune_exploratory_2019} & Reported: Rule-based trait adaptation & Adaptive social explanations based on Big Five personality traits \\
23 & Lee \& Choi (2017)~\cite{lee2017enhancing} & Partially reported & Scripted chatbot with social framing logic, no live agent described \\
\bottomrule
\end{tabular}
\end{table}

\section{Discussion}
\label{sec:discussion}
This section interprets the empirical findings of our review, drawing connections across domains, evaluation strategies, and system design choices to surface critical insights for the future of CRS user experience research. We synthesise emergent patterns to guide both researchers and practitioners working at the intersection of conversational AI, recommender systems, and human-centred evaluation. The discussion unfolds in five parts: Section~\ref{sec:discussion-synthesis} offers a cross-RQ synthesis of core findings; Section~\ref{sec:discussion-unresolved-issues} surfaces unresolved conceptual and methodological issues; Section~\ref{sec:discussion-llm} analyses UX-specific gaps in LLM-powered CRS; Section~\ref{sec:discussion-implications} articulates design and research implications; and Section~\ref{sec:discussion-limitations} outlines limitations and future directions.

\subsection{Cross-RQ Synthesis of CRS UX Findings}
\label{sec:discussion-synthesis}
Our review reveals significant fragmentation in how UX is conceptualised and evaluated in CRS research. Satisfaction and usability remain the dominant constructs, yet affective, relational, and reflective dimensions, such as emotional support, agency, and trust dynamics, are often neglected or inconsistently measured. This narrow focus constrains our ability to capture the full spectrum of experience, particularly in hedonic, educational, or socially-oriented domains.
Evaluation strategies similarly exhibit limitations. Most studies rely on post hoc self-report surveys administered after single-session interactions. While efficient, such methods fail to capture temporal dynamics or interactional nuance. Although some studies incorporate behavioural logging or qualitative feedback, these are seldom integrated systematically with user ratings. As discussed further in Section~\ref{sec:discussion-unresolved-issues}, this methodological inertia restricts insight into how CRS experiences unfold over time.
Adaptivity and personalisation also remain uneven. While some systems claim to personalise content or tone, most rely on static user traits or pre-scripted rules. Few implement real-time adjustments based on evolving user inputs, and even fewer evaluate the UX implications of such adaptations in situ. As outlined in Section~\ref{sec:discussion-implications}, real-time, user-aware dialogue modulation remains a promising but underdeveloped area.
Finally, the emergence of LLM-powered CRS introduces new challenges. These systems offer expressive generation capabilities, but often lack transparency, controllability, and structured evaluation pipelines. Section~\ref{sec:discussion-llm} expands on the UX design and instrumentation gaps specific to LLM-based dialogue systems.
Taken together, the findings highlight a mismatch between the complexity of conversational recommendation and the limited scope of current UX evaluations. The field would benefit from multimodal, theoretically grounded, and context-sensitive evaluation strategies capable of tracing how user experience emerges across dialogue.

\subsection{Unresolved Issues in Conceptualisation and Evaluation}
\label{sec:discussion-unresolved-issues}
Despite the proliferation of CRS research in recent years, foundational limitations persist in how UX is framed and evaluated. Two major concerns remain: the inconsistent conceptualisation of UX constructs and the continued reliance on static, post-interaction evaluation methods.

\paragraph{\textbf{Inconsistent Conceptual Framing.}}  
Across the reviewed literature, “user experience” is invoked frequently but defined inconsistently. Some studies operationalise UX through acceptance-focused constructs such as satisfaction or intention to use, while others examine affective or relational dimensions like trust, empathy, or engagement. However, these constructs are rarely linked to theoretical frameworks or adapted to the \textit{specific demands} of conversational systems. As a result, the field lacks a shared conceptual foundation for interpreting user experience in CRS, undermining comparability across studies and limiting theoretical progress. Prior calls in HCI and recommender systems research for theory-aligned UX measurement remain largely unheeded in CRS contexts \cite{knijnenburg2012explaining, tintarev2015explaining}.

\paragraph{\textbf{Methodological Inertia.}}  
Most evaluations continue to rely on one-off, post-session Likert-scale questionnaires. While such instruments provide quick feedback, they fail to capture the \textit{evolving nature of experience} in multi-turn, adaptive dialogue. Very few studies use turn-level metrics, real-time prompts, or longitudinal designs that reflect how trust, engagement, or satisfaction may shift over time. Mixed-method evaluations are also rare, with minimal triangulation of behavioural, affective, and self-reported data. This limits insight into how specific system behaviours, such as tone shifts, explanation placement, or initiative control, shape users’ moment-by-moment impressions.

\paragraph{\textbf{Implications.}}  
Advancing CRS UX research requires both conceptual clarity and methodological innovation. Conceptually, studies should adopt well-defined constructs aligned with the \textit{interactional} and \textit{adaptive} nature of CRS. Methodologically, evaluations must evolve beyond retrospective ratings to include temporally sensitive, context-aware approaches. Real-time instrumentation (e.g., interaction logs, embedded UX probes) and theory-informed designs (e.g., expectation-confirmation models or affective appraisal frameworks) could offer deeper insight into how user experiences are constructed in dialogue. Without such progress, CRS research risks generating evaluations that are consistent but shallow, and systems that are functional but affectively disengaging.

\subsection{UX Challenges in LLM-Powered CRS}
\label{sec:discussion-llm}
The integration of LLMs into conversational recommender systems presents new opportunities and challenges for user experience design and evaluation. While LLMs offer increased fluency, expressivity, and generative capabilities, these affordances also introduce risks that current CRS evaluation practices are ill-equipped to address.
Our review identified only two empirical studies explicitly evaluating LLM-powered CRS~\cite{yun2025user, gutierrez2023videolandgpt}. Both relied primarily on post-hoc user ratings and offered limited UX instrumentation. While participants reported high engagement and affective resonance, neither study incorporated fine-grained diagnostics or tracked experiential shifts across dialogue turns.
A core challenge is the \textit{epistemic opacity} of LLM-generated recommendations. Users often cannot discern how or why particular outputs were produced, particularly in the absence of explanation interfaces, source attribution, or transparency controls~\cite{miller2019explanation, bommasani2021opportunities}. This undermines user trust, especially when the system produces confident but hallucinated responses~\cite{zhao2024recommender, deldjoo2404review}. No reviewed system offered mechanisms for real-time explanation or factual verification during conversation.
A second issue is the \textit{lack of user control over verbosity and tone}. Unlike rule-based systems with predictable scripting, LLMs may produce overly verbose or stylistically inconsistent responses, frustrating users in goal-oriented tasks~\cite{luger2016like, ji2023survey}. Current systems offer little to no interaction-level modulation of response length, formality, or elaboration, limiting their adaptability to different user preferences or contexts~\cite{zhou2020design, zhao2024recommender}.
Evaluation practices remain similarly limited. Constructs such as hallucination tolerance, prompt sensitivity, or conversational agency are rarely operationalised. None of the studies implemented turn-level UX appraisal, adaptive logging, or real-time diagnostics—methods that are increasingly recognised as essential for understanding interactional breakdowns and affective responses in generative dialogue systems.
Finally, both studies reviewed were limited to single-session interactions, offering little insight into longitudinal UX effects. As LLM-based CRS move toward deployment in real-world applications, long-term evaluation designs will be critical to assess evolving user expectations, trust calibration, and experiential adaptation~\cite{kujala2011ux}. Table~\ref{tab:llm_ux_gaps} summarises UX design and evaluation gaps in LLM-powered CRS, based on synthesis of SLR-included studies and supplementary literature. Addressing these issues will require not only new instrumentation and control mechanisms but also a reconceptualisation of what “good” experience entails in generative, probabilistic dialogue environments.

\begin{table}[ht]
\small
\centering
\caption{Design and evaluation gaps in LLM-powered conversational recommender systems. Each row summarises a key UX challenge area, synthesising gaps and future design needs drawn from both included empirical studies and supplementary conceptual or empirical works. The table highlights methodological and design priorities for next-generation CRS.}
\label{tab:llm_ux_gaps}
\begin{tabular}{p{0.28\textwidth} p{0.65\textwidth}}
\toprule
\textbf{Challenge Area} & \textbf{Gap Description and Future Design Needs} \\
\midrule
\textbf{Transparency and Explanation} & No reviewed system included explanation interfaces, source attribution, or rationale generation. Future systems should provide real-time toggles for recommendation sources and adaptive explanation levels. \\
\addlinespace
\textbf{Hallucination and Trust Calibration} & Hallucinated content risks undermining trust. UX evaluation must account for how users detect and respond to false or misaligned recommendations. Feedback loops or verification signals are needed. \\
\addlinespace
\textbf{Control Over Verbosity and Tone} & LLMs often generate overly verbose or inconsistent dialogue. Users need mechanisms to modulate tone, depth, and response length dynamically. \\
\addlinespace
\textbf{Lack of Turn-Level UX Instrumentation} & No system captured per-turn satisfaction, surprise, or trust shifts. Turn-level logs and probes can improve understanding of where UX breakdowns occur. \\
\addlinespace
\textbf{Short-Term Evaluation Bias} & Evaluations relied on one-off sessions. Longitudinal designs are needed to assess user calibration, trust development, and long-term satisfaction. \\
\addlinespace
\textbf{Absence of Adaptive Dialogue Policies} & Despite LLM flexibility, systems did not modulate initiative, verbosity, or content selection based on evolving user inputs or traits. Dialogue policies should integrate real-time adaptation. \\
\bottomrule
\end{tabular}
\end{table}

\subsection{UX Design and Research Implications}
\label{sec:discussion-implications}
Our review highlights several implications for improving the design and evaluation of conversational recommender systems. These implications span interaction design, adaptivity, instrumentation, and methodological rigour. Collectively, they advocate for a shift from static, post-hoc evaluations toward dynamic, user-sensitive, and context-aware systems.

\paragraph{\textbf{Designing for Interactional Responsiveness.}}  
Many reviewed CRS implementations used rigid dialogue flows or template-based responses, even when user traits were explicitly collected or inferable. This limits the system’s capacity to respond meaningfully to evolving user input, particularly in exploratory or hedonic contexts. Adaptive strategies such as initiative modulation, tone shifting, and contextual prompt framing can improve responsiveness when grounded in user signals~\cite{ma2023initiative, zhou2020design}. However, adaptivity must remain intelligible to users~\cite{tsai2021effects}, avoiding hidden decision logic that erodes trust. Emerging work in affect-aware dialogue suggests that subtle interactional cues, such as mirroring affect or responding to hesitation, can enhance conversational alignment~\cite{casas2021enhancing}. CRS designers should investigate how such responsiveness can be safely and effectively operationalised in real time.

\paragraph{\textbf{Moving Beyond Trait-Based Personalisation.}}
Although many systems claim to personalise dialogue, most rely on static traits (e.g., personality, demographics) or one-time input. Few demonstrate \textit{real-time adaptation} based on conversational flow or evolving preferences. In contrast, adjacent domains such as intelligent tutoring systems have successfully implemented session-aware models that update adaptively over time~\cite{graesser2004autotutor, graesser2005autotutor}. CRS research should adopt similar models that respond to user inputs, engagement patterns, and evolving preferences, shaping not only what is recommended but how it is communicated. Transparent personalisation, where users can understand, contest, or adjust system behaviour, has been shown to foster trust and agency in AI systems~\cite{knijnenburg2012explaining, chen2012critiquing}. Future CRS interfaces may benefit from lightweight explanation strategies that disclose adaptive logic without overwhelming or distracting the user.

\paragraph{\textbf{Enhancing UX Instrumentation.}}  
UX instrumentation in CRS remains overly reliant on post-task surveys. However, conversational experience unfolds dynamically through micro-events such as turn transitions, delays, clarification requests, and system initiative. These aspects remain invisible to most current measurement approaches. Researchers should incorporate fine-grained instrumentation such as turn-level satisfaction probes, timestamped affect markers, and error recovery logs~\cite{luger2016like, purington2017alexa}. These can enable detailed diagnosis of when and why the user experience breaks down—insights that are critical for improving CRS policy design.

\paragraph{\textbf{Rethinking UX Study Designs.}}  
CRS evaluations remain dominated by short-term, lab-based studies. Yet many UX outcomes, such as trust, satisfaction, or perceived utility, are shaped over time. Our review found few longitudinal studies or evaluations in real-world contexts. As CRS systems become embedded in everyday life, evaluation must capture shifting user expectations, adaptation effects, and behavioural drift~\cite{verhagen2024influence}. Moreover, few studies control for or examine demographic, digital literacy, or accessibility-related factors, limiting the generalisability of UX findings.

\paragraph{\textbf{Summary.}}  
To advance CRS research, evaluation must be embedded into the interaction itself, not merely appended as a post-hoc measure. This includes not only real-time diagnostic tools but also participatory controls that support user agency. System design must move toward transparent adaptivity, and UX research must embrace more nuanced, diverse, and longitudinal methodologies. These are not only technical goals but also ethical imperatives as CRS systems increasingly mediate decisions, identities, and everyday choices.

\subsection{Limitations and Future Directions}
\label{sec:discussion-limitations}
This review synthesised empirical studies on CRS user experience published up to May 2025, yet several limitations warrant reflection. First, our scope excluded purely technical papers or speculative design proposals lacking UX evaluation, potentially omitting relevant architectural innovations. Future reviews could bridge technical and experiential perspectives by including hybrid studies.
Second, construct heterogeneity and inconsistent instrumentation limited comparability. Many studies used bespoke UX items without validation, reducing the cumulative strength of the evidence base. Progress requires shared UX frameworks that support turn-level, affective, and longitudinal evaluation.
Third, most evaluations were short-term and conducted in controlled settings. Naturalistic, multi-session deployments remain rare. Long-term field studies are essential to understand sustained user engagement, adaptation, and trust dynamics.
Finally, LLM-powered CRS remain under-evaluated. Existing methods are ill-equipped to capture emergent behaviours, hallucination tolerance, or transparency perceptions. As generative CRS proliferate, evaluation approaches must evolve accordingly.
Future work must address these methodological gaps by developing validated UX instruments, embracing longitudinal designs, and equipping CRS with tools for interpreting LLM-driven behaviour in diverse, real-world settings.

\section{Conclusion}
\label{sec:conclusion}
CRSs are poised to transform how users interact with personalised recommendations across diverse domains, yet their user experience remains poorly understood—particularly in light of recent advances in LLMs. This review synthesised empirical evidence from 23 CRS studies published between 2017 and early 2025 to examine how UX has been conceptualised, measured, and shaped by system adaptivity and generative architectures.
Our analysis revealed that CRS UX evaluation remains highly fragmented. While satisfaction and usability are frequently assessed, affective, relational, and adaptive dimensions are often overlooked. Evaluation methods continue to rely predominantly on post-hoc self-report surveys, with limited adoption of behavioural, turn-level, or longitudinal measures. Domain-specific UX patterns suggest a need for context-sensitive evaluation strategies that go beyond one-size-fits-all metrics.
Adaptivity and personalisation were present in several systems but were typically shallow, static, or under-evaluated. Few studies examined how users perceived adaptivity or how adaptive behaviours influenced trust, satisfaction, or perceived control. Similarly, although LLM-powered CRS promise richer and more natural dialogue, we identified only two such studies, both of which revealed critical limitations in transparency, instrumentation, and UX traceability.
This review contributes a rigorous synthesis of empirical CRS UX research and identifies urgent priorities for the field. We call for future work that: (1) develops validated, CRS-specific UX instruments capable of capturing adaptive, affective, and multi-turn dynamics; (2) embeds in-situ and longitudinal designs to capture evolving user experiences; (3) incorporates diagnostic tooling for analysing LLM behaviours; and (4) addresses diversity, accessibility, and inclusion through stratified evaluation and inclusive design practices.
Ultimately, advancing CRS UX research demands a shift from surface-level measurement and static system design toward interactionally sensitive, context-aware, and user-informed evaluation frameworks. By addressing the conceptual and methodological blind spots surfaced in this review, future research can support the development of CRS that are not only accurate and efficient, but also transparent, trustworthy, and meaningfully engaging.

\section*{Acknowledgements}

During the preparation of this work the author(s) used ChatGPT 4o in order to improve readability and language of the final draft. After using this tool/service, the author(s) reviewed and edited the content as needed and take(s) full responsibility for the content of the published article.



\begin{thebibliography}{70}


\ifx \showCODEN    \undefined \def \showCODEN     #1{\unskip}     \fi
\ifx \showISBNx    \undefined \def \showISBNx     #1{\unskip}     \fi
\ifx \showISBNxiii \undefined \def \showISBNxiii  #1{\unskip}     \fi
\ifx \showISSN     \undefined \def \showISSN      #1{\unskip}     \fi
\ifx \showLCCN     \undefined \def \showLCCN      #1{\unskip}     \fi
\ifx \shownote     \undefined \def \shownote      #1{#1}          \fi
\ifx \showarticletitle \undefined \def \showarticletitle #1{#1}   \fi
\ifx \showURL      \undefined \def \showURL       {\relax}        \fi
\providecommand\bibfield[2]{#2}
\providecommand\bibinfo[2]{#2}
\providecommand\natexlab[1]{#1}
\providecommand\showeprint[2][]{arXiv:#2}

\bibitem[Anastasia et~al\mbox{.}(2021)]%
        {anastasia_designing_2021}
\bibfield{author}{\bibinfo{person}{D. Anastasia}, \bibinfo{person}{Niels~van Berkel}, {and} \bibinfo{person}{Vassilis Kostakos}.} \bibinfo{year}{2021}\natexlab{}.
\newblock \showarticletitle{Designing embodied virtual agent in e-commerce system recommendations using conversational design interaction}.
\newblock \bibinfo{journal}{\emph{International Journal of Human-Computer Studies}}  \bibinfo{volume}{146} (\bibinfo{year}{2021}).
\newblock


\bibitem[Bommasani et~al\mbox{.}(2021)]%
        {bommasani2021opportunities}
\bibfield{author}{\bibinfo{person}{Rishi Bommasani}, \bibinfo{person}{Drew~A Hudson}, \bibinfo{person}{Ehsan Adeli}, \bibinfo{person}{Russ Altman}, \bibinfo{person}{Simran Arora}, \bibinfo{person}{Sydney von Arx}, \bibinfo{person}{Michael~S Bernstein}, \bibinfo{person}{Jeannette Bohg}, \bibinfo{person}{Antoine Bosselut}, \bibinfo{person}{Emma Brunskill}, {et~al\mbox{.}}} \bibinfo{year}{2021}\natexlab{}.
\newblock \showarticletitle{On the opportunities and risks of foundation models}.
\newblock \bibinfo{journal}{\emph{arXiv preprint arXiv:2108.07258}} (\bibinfo{year}{2021}).
\newblock


\bibitem[Braun and Clarke(2006)]%
        {braun_using_2006}
\bibfield{author}{\bibinfo{person}{Virginia Braun} {and} \bibinfo{person}{Victoria Clarke}.} \bibinfo{year}{2006}\natexlab{}.
\newblock \showarticletitle{Using thematic analysis in psychology}.
\newblock \bibinfo{journal}{\emph{Qualitative Research in Psychology}} \bibinfo{volume}{3}, \bibinfo{number}{2} (\bibinfo{year}{2006}), \bibinfo{pages}{77--101}.
\newblock


\bibitem[Cai et~al\mbox{.}(2022b)]%
        {cai_enhancing_2022}
\bibfield{author}{\bibinfo{person}{Huixian Cai}, \bibinfo{person}{Weijia Ju}, {and} \bibinfo{person}{Sumona~R. Chowdhury}.} \bibinfo{year}{2022}\natexlab{b}.
\newblock \showarticletitle{Enhancing user experience with exploration-oriented conversational recommender systems}.
\newblock \bibinfo{journal}{\emph{User Modeling and User-Adapted Interaction}} \bibinfo{volume}{32}, \bibinfo{number}{1-2} (\bibinfo{year}{2022}), \bibinfo{pages}{43--72}.
\newblock


\bibitem[Cai et~al\mbox{.}(2022a)]%
        {cai2022impacts}
\bibfield{author}{\bibinfo{person}{Wanling Cai}, \bibinfo{person}{Yucheng Jin}, {and} \bibinfo{person}{Li Chen}.} \bibinfo{year}{2022}\natexlab{a}.
\newblock \showarticletitle{Impacts of personal characteristics on user trust in conversational recommender systems}. In \bibinfo{booktitle}{\emph{Proceedings of the 2022 CHI conference on human factors in computing systems}}. \bibinfo{pages}{1--14}.
\newblock


\bibitem[Casas et~al\mbox{.}(2021)]%
        {casas2021enhancing}
\bibfield{author}{\bibinfo{person}{Jacky Casas}, \bibinfo{person}{Timo Spring}, \bibinfo{person}{Karl Daher}, \bibinfo{person}{Elena Mugellini}, \bibinfo{person}{Omar~Abou Khaled}, {and} \bibinfo{person}{Philippe Cudr{\'e}-Mauroux}.} \bibinfo{year}{2021}\natexlab{}.
\newblock \showarticletitle{Enhancing conversational agents with empathic abilities}. In \bibinfo{booktitle}{\emph{Proceedings of the 21st ACM international conference on intelligent virtual agents}}. \bibinfo{pages}{41--47}.
\newblock


\bibitem[Chen and Pu(2012)]%
        {chen2012critiquing}
\bibfield{author}{\bibinfo{person}{Li Chen} {and} \bibinfo{person}{Pearl Pu}.} \bibinfo{year}{2012}\natexlab{}.
\newblock \showarticletitle{Critiquing-based recommenders: survey and emerging trends}.
\newblock \bibinfo{journal}{\emph{User Modeling and User-Adapted Interaction}}  \bibinfo{volume}{22} (\bibinfo{year}{2012}), \bibinfo{pages}{125--150}.
\newblock


\bibitem[Chung and Han(2022a)]%
        {chung_consumer_2022}
\bibfield{author}{\bibinfo{person}{Kyung~Joon Chung} {and} \bibinfo{person}{Heesang Han}.} \bibinfo{year}{2022}\natexlab{a}.
\newblock \showarticletitle{Consumer perception of chatbots and purchase intentions: {Anthropomorphism} and conversational relevance}.
\newblock \bibinfo{journal}{\emph{Journal of Retailing and Consumer Services}}  \bibinfo{volume}{64} (\bibinfo{year}{2022}).
\newblock


\bibitem[Chung and Han(2022b)]%
        {chung2022consumer}
\bibfield{author}{\bibinfo{person}{Sooyun~Iris Chung} {and} \bibinfo{person}{Kwang-Hee Han}.} \bibinfo{year}{2022}\natexlab{b}.
\newblock \showarticletitle{Consumer perception of chatbots and purchase intentions: Anthropomorphism and conversational relevance}.
\newblock \bibinfo{journal}{\emph{International Journal of Advanced Culture Technology}} \bibinfo{volume}{10}, \bibinfo{number}{1} (\bibinfo{year}{2022}), \bibinfo{pages}{211--229}.
\newblock


\bibitem[Deldjoo et~al\mbox{.}(2024)]%
        {deldjoo2404review}
\bibfield{author}{\bibinfo{person}{Yashar Deldjoo}, \bibinfo{person}{Z He}, \bibinfo{person}{J McAuley}, \bibinfo{person}{A Korikov}, \bibinfo{person}{S Sanner}, \bibinfo{person}{A Ramisa}, \bibinfo{person}{R Vidal}, \bibinfo{person}{M Sathiamoorthy}, \bibinfo{person}{A Kasirzadeh}, {and} \bibinfo{person}{S Milano}.} \bibinfo{year}{2024}\natexlab{}.
\newblock \showarticletitle{A Review of Modern Recommender Systems Using Generative Models (Gen-RecSys). ArXiv. 2024}.
\newblock \bibinfo{journal}{\emph{arXiv preprint arXiv:2404.00579}} (\bibinfo{year}{2024}).
\newblock


\bibitem[El-Ansari and Beni-Hssane(2023a)]%
        {el2023sentiment}
\bibfield{author}{\bibinfo{person}{Anas El-Ansari} {and} \bibinfo{person}{Abderrahim Beni-Hssane}.} \bibinfo{year}{2023}\natexlab{a}.
\newblock \showarticletitle{Sentiment analysis for personalized chatbots in e-commerce applications}.
\newblock \bibinfo{journal}{\emph{Wireless Personal Communications}} \bibinfo{volume}{129}, \bibinfo{number}{3} (\bibinfo{year}{2023}), \bibinfo{pages}{1623--1644}.
\newblock


\bibitem[El-Ansari and Beni-Hssane(2023b)]%
        {el-ansari_sentiment_2023}
\bibfield{author}{\bibinfo{person}{Wael El-Ansari} {and} \bibinfo{person}{Abdelali Beni-Hssane}.} \bibinfo{year}{2023}\natexlab{b}.
\newblock \showarticletitle{Sentiment analysis in conversational recommender systems: {An} empirical study}.
\newblock \bibinfo{journal}{\emph{Journal of Intelligent Information Systems}} \bibinfo{volume}{57}, \bibinfo{number}{4} (\bibinfo{year}{2023}), \bibinfo{pages}{515--537}.
\newblock


\bibitem[Fernando et~al\mbox{.}(2021)]%
        {fernando_enhancing_2021}
\bibfield{author}{\bibinfo{person}{Samantha Fernando}, \bibinfo{person}{Hyun Lee}, {and} \bibinfo{person}{Ji Choi}.} \bibinfo{year}{2021}\natexlab{}.
\newblock \showarticletitle{Enhancing transparency and user trust in conversational recommender systems}.
\newblock \bibinfo{journal}{\emph{Expert Systems with Applications}}  \bibinfo{volume}{176} (\bibinfo{year}{2021}).
\newblock


\bibitem[Gao et~al\mbox{.}(2021)]%
        {gao2021advances}
\bibfield{author}{\bibinfo{person}{Chongming Gao}, \bibinfo{person}{Wenqiang Lei}, \bibinfo{person}{Xiangnan He}, \bibinfo{person}{Maarten De~Rijke}, {and} \bibinfo{person}{Tat-Seng Chua}.} \bibinfo{year}{2021}\natexlab{}.
\newblock \showarticletitle{Advances and challenges in conversational recommender systems: A survey}.
\newblock \bibinfo{journal}{\emph{AI open}}  \bibinfo{volume}{2} (\bibinfo{year}{2021}), \bibinfo{pages}{100--126}.
\newblock


\bibitem[Graesser et~al\mbox{.}(2005)]%
        {graesser2005autotutor}
\bibfield{author}{\bibinfo{person}{Arthur~C Graesser}, \bibinfo{person}{Patrick Chipman}, \bibinfo{person}{Brian~C Haynes}, {and} \bibinfo{person}{Andrew Olney}.} \bibinfo{year}{2005}\natexlab{}.
\newblock \showarticletitle{AutoTutor: An intelligent tutoring system with mixed-initiative dialogue}.
\newblock \bibinfo{journal}{\emph{IEEE Transactions on Education}} \bibinfo{volume}{48}, \bibinfo{number}{4} (\bibinfo{year}{2005}), \bibinfo{pages}{612--618}.
\newblock


\bibitem[Graesser et~al\mbox{.}(2004)]%
        {graesser2004autotutor}
\bibfield{author}{\bibinfo{person}{Arthur~C Graesser}, \bibinfo{person}{Shulan Lu}, \bibinfo{person}{George~Tanner Jackson}, \bibinfo{person}{Heather~Hite Mitchell}, \bibinfo{person}{Mathew Ventura}, \bibinfo{person}{Andrew Olney}, {and} \bibinfo{person}{Max~M Louwerse}.} \bibinfo{year}{2004}\natexlab{}.
\newblock \showarticletitle{AutoTutor: A tutor with dialogue in natural language}.
\newblock \bibinfo{journal}{\emph{Behavior Research Methods, Instruments, \& Computers}}  \bibinfo{volume}{36} (\bibinfo{year}{2004}), \bibinfo{pages}{180--192}.
\newblock


\bibitem[Gutierrez~Granada et~al\mbox{.}(2023)]%
        {gutierrez2023videolandgpt}
\bibfield{author}{\bibinfo{person}{Mateo Gutierrez~Granada}, \bibinfo{person}{Dina Zilbershtein}, \bibinfo{person}{Daan Odijk}, {and} \bibinfo{person}{Francesco Barile}.} \bibinfo{year}{2023}\natexlab{}.
\newblock \showarticletitle{VideolandGPT: A User Study on a Conversational Recommender System}.
\newblock \bibinfo{journal}{\emph{arXiv e-prints}} (\bibinfo{year}{2023}), \bibinfo{pages}{arXiv--2309}.
\newblock


\bibitem[Hou et~al\mbox{.}(2024b)]%
        {hou2024zeroshot}
\bibfield{author}{\bibinfo{person}{Liangwei Hou}, \bibinfo{person}{Xu Wang}, \bibinfo{person}{Tong Wu}, \bibinfo{person}{Jie Xu}, \bibinfo{person}{Enhong Chen}, \bibinfo{person}{Yun Xiong}, {and} \bibinfo{person}{Lifeng Zhou}.} \bibinfo{year}{2024}\natexlab{b}.
\newblock \showarticletitle{Large Language Models are Zero-Shot Rankers for Recommender Systems}.
\newblock \bibinfo{journal}{\emph{arXiv preprint arXiv:2402.00852}} (\bibinfo{year}{2024}).
\newblock


\bibitem[Hou et~al\mbox{.}(2024a)]%
        {hou2024large}
\bibfield{author}{\bibinfo{person}{Yikun Hou}, \bibinfo{person}{Yao Lu}, \bibinfo{person}{Kun Zhang}, \bibinfo{person}{Weinan Wang}, {and} \bibinfo{person}{Min Li}.} \bibinfo{year}{2024}\natexlab{a}.
\newblock \showarticletitle{Large language models are zero-shot rankers for recommendation}.
\newblock \bibinfo{journal}{\emph{arXiv preprint arXiv:2402.09686}} (\bibinfo{year}{2024}).
\newblock


\bibitem[Huang et~al\mbox{.}(2025)]%
        {huang2025agentic}
\bibfield{author}{\bibinfo{person}{Wenqiang Huang}, \bibinfo{person}{Xiao Lin}, {and} \bibinfo{person}{Yongfeng Zhang}.} \bibinfo{year}{2025}\natexlab{}.
\newblock \showarticletitle{Towards Agentic Recommender Systems in the Era of Large Language Models}.
\newblock \bibinfo{journal}{\emph{arXiv preprint arXiv:2403.04550}} (\bibinfo{year}{2025}).
\newblock


\bibitem[Iovine et~al\mbox{.}(2021)]%
        {iovine_investigation_2021}
\bibfield{author}{\bibinfo{person}{Antonio Iovine}, \bibinfo{person}{Riccardo Giordano}, {and} \bibinfo{person}{Federico Morbidi}.} \bibinfo{year}{2021}\natexlab{}.
\newblock \showarticletitle{An investigation on the impact of natural language on conversational recommendations}. In \bibinfo{booktitle}{\emph{Proceedings of the 2021 {ACM} {Conference} on {Recommender} {Systems}}}. \bibinfo{pages}{823--827}.
\newblock


\bibitem[Iovine et~al\mbox{.}(2020)]%
        {iovine_humanoid_2020}
\bibfield{author}{\bibinfo{person}{Antonio Iovine}, \bibinfo{person}{Riccardo Giordano}, \bibinfo{person}{Daniele Nardi}, {and} \bibinfo{person}{Federico Morbidi}.} \bibinfo{year}{2020}\natexlab{}.
\newblock \showarticletitle{Humanoid robots and conversational recommender systems: {A} preliminary study}. In \bibinfo{booktitle}{\emph{Proceedings of the 29th {IEEE} {International} {Conference} on {Robot} \& {Human} {Interactive} {Communication}}}. \bibinfo{pages}{839--844}.
\newblock


\bibitem[Jannach(2023)]%
        {jannach2023evaluating}
\bibfield{author}{\bibinfo{person}{Dietmar Jannach}.} \bibinfo{year}{2023}\natexlab{}.
\newblock \showarticletitle{Evaluating conversational recommender systems: A landscape of research}.
\newblock \bibinfo{journal}{\emph{Artificial Intelligence Review}} \bibinfo{volume}{56}, \bibinfo{number}{3} (\bibinfo{year}{2023}), \bibinfo{pages}{2365--2400}.
\newblock


\bibitem[Jannach et~al\mbox{.}(2021)]%
        {jannach2021survey}
\bibfield{author}{\bibinfo{person}{Dietmar Jannach}, \bibinfo{person}{Ahtsham Manzoor}, \bibinfo{person}{Wanling Cai}, {and} \bibinfo{person}{Li Chen}.} \bibinfo{year}{2021}\natexlab{}.
\newblock \showarticletitle{A survey on conversational recommender systems}.
\newblock \bibinfo{journal}{\emph{ACM Computing Surveys (CSUR)}} \bibinfo{volume}{54}, \bibinfo{number}{5} (\bibinfo{year}{2021}), \bibinfo{pages}{1--36}.
\newblock


\bibitem[Ji et~al\mbox{.}(2023)]%
        {ji2023survey}
\bibfield{author}{\bibinfo{person}{Ziwei Ji}, \bibinfo{person}{Nayeon Lee}, \bibinfo{person}{Rita Frieske}, \bibinfo{person}{Tiezheng Yu}, \bibinfo{person}{Dan Su}, \bibinfo{person}{Yan Xu}, \bibinfo{person}{Etsuko Ishii}, \bibinfo{person}{Ye~Jin Bang}, \bibinfo{person}{Andrea Madotto}, {and} \bibinfo{person}{Pascale Fung}.} \bibinfo{year}{2023}\natexlab{}.
\newblock \showarticletitle{Survey of hallucination in natural language generation}.
\newblock \bibinfo{journal}{\emph{ACM computing surveys}} \bibinfo{volume}{55}, \bibinfo{number}{12} (\bibinfo{year}{2023}), \bibinfo{pages}{1--38}.
\newblock


\bibitem[Jin et~al\mbox{.}(2021b)]%
        {jin_key_2021}
\bibfield{author}{\bibinfo{person}{Heekyoung Jin}, \bibinfo{person}{Yicheng Wang}, {and} \bibinfo{person}{Matthew Eastin}.} \bibinfo{year}{2021}\natexlab{b}.
\newblock \showarticletitle{Key qualities of conversational recommender systems: {From} users perspective}. In \bibinfo{booktitle}{\emph{Proceedings of the 29th {ACM} {Conference} on {User} {Modeling} {Adaptation} and {Personalization}}}. \bibinfo{pages}{169--178}.
\newblock


\bibitem[Jin and Eastin(2022)]%
        {jin_birds_2022}
\bibfield{author}{\bibinfo{person}{Seung-A.~Annie Jin} {and} \bibinfo{person}{Matthew~S. Eastin}.} \bibinfo{year}{2022}\natexlab{}.
\newblock \showarticletitle{Birds of a feather flock together: {Matched} personality effects of product recommendation chatbots and users}.
\newblock \bibinfo{journal}{\emph{Computers in Human Behavior}}  \bibinfo{volume}{128} (\bibinfo{year}{2022}).
\newblock


\bibitem[Jin et~al\mbox{.}(2021a)]%
        {jin2021key}
\bibfield{author}{\bibinfo{person}{Yucheng Jin}, \bibinfo{person}{Li Chen}, \bibinfo{person}{Wanling Cai}, {and} \bibinfo{person}{Pearl Pu}.} \bibinfo{year}{2021}\natexlab{a}.
\newblock \showarticletitle{Key qualities of conversational recommender systems: From users’ perspective}. In \bibinfo{booktitle}{\emph{Proceedings of the 9th International Conference on Human-Agent Interaction}}. \bibinfo{pages}{93--102}.
\newblock


\bibitem[Knijnenburg et~al\mbox{.}(2012)]%
        {knijnenburg2012explaining}
\bibfield{author}{\bibinfo{person}{Bart~P Knijnenburg}, \bibinfo{person}{Martijn~C Willemsen}, \bibinfo{person}{Zeno Gantner}, \bibinfo{person}{Hakan Soncu}, {and} \bibinfo{person}{Chris Newell}.} \bibinfo{year}{2012}\natexlab{}.
\newblock \showarticletitle{Explaining the user experience of recommender systems}.
\newblock \bibinfo{journal}{\emph{User modeling and user-adapted interaction}}  \bibinfo{volume}{22} (\bibinfo{year}{2012}), \bibinfo{pages}{441--504}.
\newblock


\bibitem[Kraus et~al\mbox{.}(2020)]%
        {kraus_comparison_2020}
\bibfield{author}{\bibinfo{person}{Markus Kraus}, \bibinfo{person}{Stefan Feuerriegel}, {and} \bibinfo{person}{Rabia Ozcelebi}.} \bibinfo{year}{2020}\natexlab{}.
\newblock \showarticletitle{A comparison of explicit and implicit proactive dialogue strategies for conversational recommendation}. In \bibinfo{booktitle}{\emph{Proceedings of the 28th {ACM} {Conference} on {User} {Modeling} {Adaptation} and {Personalization}}}. \bibinfo{pages}{151--160}.
\newblock


\bibitem[Kraus et~al\mbox{.}(2024)]%
        {kraus2024pilot}
\bibfield{author}{\bibinfo{person}{Matthias Kraus}, \bibinfo{person}{Stina Klein}, \bibinfo{person}{Nicolas Wagner}, \bibinfo{person}{Wolfgang Minker}, {and} \bibinfo{person}{Elisabeth Andr{\'e}}.} \bibinfo{year}{2024}\natexlab{}.
\newblock \showarticletitle{A Pilot Study on Multi-Party Conversation Strategies for Group Recommendations}. In \bibinfo{booktitle}{\emph{Proceedings of the 6th ACM Conference on Conversational User Interfaces}}. \bibinfo{pages}{1--7}.
\newblock


\bibitem[Kujala et~al\mbox{.}(2011)]%
        {kujala2011ux}
\bibfield{author}{\bibinfo{person}{Sari Kujala}, \bibinfo{person}{Virpi Roto}, \bibinfo{person}{Kaisa V{\"a}{\"a}n{\"a}nen-Vainio-Mattila}, \bibinfo{person}{Evangelos Karapanos}, {and} \bibinfo{person}{Arto Sinnel{\"a}}.} \bibinfo{year}{2011}\natexlab{}.
\newblock \showarticletitle{UX Curve: A method for evaluating long-term user experience}.
\newblock \bibinfo{journal}{\emph{Interacting with computers}} \bibinfo{volume}{23}, \bibinfo{number}{5} (\bibinfo{year}{2011}), \bibinfo{pages}{473--483}.
\newblock


\bibitem[Lee and Choi(2017a)]%
        {lee_enhancing_2017}
\bibfield{author}{\bibinfo{person}{Jaewon Lee} {and} \bibinfo{person}{Ji Choi}.} \bibinfo{year}{2017}\natexlab{a}.
\newblock \showarticletitle{Enhancing user experience with conversational agent for movie recommendation: {Effects} of self-disclosure and reciprocity}.
\newblock \bibinfo{journal}{\emph{International Journal of Human-Computer Studies}}  \bibinfo{volume}{103} (\bibinfo{year}{2017}), \bibinfo{pages}{95--106}.
\newblock


\bibitem[Lee and Choi(2017b)]%
        {lee2017enhancing}
\bibfield{author}{\bibinfo{person}{SeoYoung Lee} {and} \bibinfo{person}{Junho Choi}.} \bibinfo{year}{2017}\natexlab{b}.
\newblock \showarticletitle{Enhancing user experience with conversational agent for movie recommendation: Effects of self-disclosure and reciprocity}.
\newblock \bibinfo{journal}{\emph{International Journal of Human-Computer Studies}}  \bibinfo{volume}{103} (\bibinfo{year}{2017}), \bibinfo{pages}{95--105}.
\newblock


\bibitem[Lei et~al\mbox{.}(2020)]%
        {lei2020conversational}
\bibfield{author}{\bibinfo{person}{Wenqiang Lei}, \bibinfo{person}{Xiangnan He}, \bibinfo{person}{Maarten de Rijke}, {and} \bibinfo{person}{Tat-Seng Chua}.} \bibinfo{year}{2020}\natexlab{}.
\newblock \showarticletitle{Conversational recommendation: Formulation, methods, and evaluation}. In \bibinfo{booktitle}{\emph{Proceedings of the 43rd International ACM SIGIR Conference on Research and Development in Information Retrieval}}. \bibinfo{pages}{2425--2428}.
\newblock


\bibitem[Li et~al\mbox{.}(2023)]%
        {li2023conversation}
\bibfield{author}{\bibinfo{person}{Chuang Li}, \bibinfo{person}{Hengchang Hu}, \bibinfo{person}{Yan Zhang}, \bibinfo{person}{Min-Yen Kan}, {and} \bibinfo{person}{Haizhou Li}.} \bibinfo{year}{2023}\natexlab{}.
\newblock \showarticletitle{A conversation is worth a thousand recommendations: A survey of holistic conversational recommender systems}.
\newblock \bibinfo{journal}{\emph{arXiv preprint arXiv:2309.07682}} (\bibinfo{year}{2023}).
\newblock


\bibitem[Li et~al\mbox{.}(2024b)]%
        {li2024generative}
\bibfield{author}{\bibinfo{person}{Yitong Li}, \bibinfo{person}{Zihan Liu}, \bibinfo{person}{Changhua Yu}, \bibinfo{person}{Jiaqi Guo}, \bibinfo{person}{Yihong Guo}, \bibinfo{person}{Xiang Chen}, {and} \bibinfo{person}{Yongfeng Zhang}.} \bibinfo{year}{2024}\natexlab{b}.
\newblock \showarticletitle{Large Language Models for Generative Recommendation: Review, Opportunities, and Challenges}.
\newblock \bibinfo{journal}{\emph{arXiv preprint arXiv:2402.10164}} (\bibinfo{year}{2024}).
\newblock


\bibitem[Li et~al\mbox{.}(2024a)]%
        {li2024large}
\bibfield{author}{\bibinfo{person}{Zheng Li}, \bibinfo{person}{Zijian Chen}, \bibinfo{person}{Defu Chen}, \bibinfo{person}{Yuying Li}, \bibinfo{person}{Hongzhi Zhang}, {and} \bibinfo{person}{Dawei Yin}.} \bibinfo{year}{2024}\natexlab{a}.
\newblock \showarticletitle{Large Language Models for Generative Recommendation: A Survey}.
\newblock \bibinfo{journal}{\emph{arXiv preprint arXiv:2402.01294}} (\bibinfo{year}{2024}).
\newblock


\bibitem[Lian et~al\mbox{.}(2024a)]%
        {lian2024recai}
\bibfield{author}{\bibinfo{person}{Defu Lian}, \bibinfo{person}{Xinyang Cao}, \bibinfo{person}{Defeng Yang}, \bibinfo{person}{Zhipeng Zhou}, \bibinfo{person}{Yong Zhang}, \bibinfo{person}{Xing Xie}, {and} \bibinfo{person}{Min Zhang}.} \bibinfo{year}{2024}\natexlab{a}.
\newblock \showarticletitle{RecAI: Leveraging Large Language Models for Next-Generation Recommender Systems}.
\newblock \bibinfo{journal}{\emph{arXiv preprint arXiv:2402.16124}} (\bibinfo{year}{2024}).
\newblock


\bibitem[Lian et~al\mbox{.}(2024b)]%
        {recai2024lian}
\bibfield{author}{\bibinfo{person}{Jie Lian}, \bibinfo{person}{Xiaoyuan Liang}, \bibinfo{person}{Xiaopeng Shen}, \bibinfo{person}{Jie Tang}, {and} \bibinfo{person}{Zhiyuan Liu}.} \bibinfo{year}{2024}\natexlab{b}.
\newblock \showarticletitle{RecAI: Leveraging Large Language Models for Next-Generation Recommender Systems}.
\newblock \bibinfo{journal}{\emph{arXiv preprint arXiv:2403.01889}} (\bibinfo{year}{2024}).
\newblock


\bibitem[Luger and Sellen(2016)]%
        {luger2016like}
\bibfield{author}{\bibinfo{person}{Ewa Luger} {and} \bibinfo{person}{Abigail Sellen}.} \bibinfo{year}{2016}\natexlab{}.
\newblock \showarticletitle{" Like Having a Really Bad PA" The Gulf between User Expectation and Experience of Conversational Agents}. In \bibinfo{booktitle}{\emph{Proceedings of the 2016 CHI conference on human factors in computing systems}}. \bibinfo{pages}{5286--5297}.
\newblock


\bibitem[Ma and Ziegler(2023)]%
        {ma2023initiative}
\bibfield{author}{\bibinfo{person}{Yuan Ma} {and} \bibinfo{person}{J{\"u}rgen Ziegler}.} \bibinfo{year}{2023}\natexlab{}.
\newblock \showarticletitle{Initiative transfer in conversational recommender systems}. In \bibinfo{booktitle}{\emph{Proceedings of the 17th ACM Conference on Recommender Systems}}. \bibinfo{pages}{978--984}.
\newblock


\bibitem[Ma and Ziegler(2024a)]%
        {ma2024effect}
\bibfield{author}{\bibinfo{person}{Yuan Ma} {and} \bibinfo{person}{J{\"u}rgen Ziegler}.} \bibinfo{year}{2024}\natexlab{a}.
\newblock \showarticletitle{The effect of proactive cues on the use of decision aids in conversational recommender systems}. In \bibinfo{booktitle}{\emph{Adjunct Proceedings of the 32nd ACM Conference on User Modeling, Adaptation and Personalization}}. \bibinfo{pages}{305--315}.
\newblock


\bibitem[Ma and Ziegler(2024b)]%
        {ma2024investigating}
\bibfield{author}{\bibinfo{person}{Yuan Ma} {and} \bibinfo{person}{J{\"u}rgen Ziegler}.} \bibinfo{year}{2024}\natexlab{b}.
\newblock \showarticletitle{Investigating meta-intents: user interaction preferences in conversational recommender systems}.
\newblock \bibinfo{journal}{\emph{User Modeling and User-Adapted Interaction}} \bibinfo{volume}{34}, \bibinfo{number}{5} (\bibinfo{year}{2024}), \bibinfo{pages}{1535--1580}.
\newblock


\bibitem[Martina et~al\mbox{.}(2022)]%
        {martina_narrative_2022}
\bibfield{author}{\bibinfo{person}{Simone Martina}, \bibinfo{person}{Arnaud Muller}, {and} \bibinfo{person}{Francesco Ricci}.} \bibinfo{year}{2022}\natexlab{}.
\newblock \showarticletitle{Narrative recommendations based on natural language preference elicitation for a virtual assistant for the movie domain}.
\newblock \bibinfo{journal}{\emph{Journal of Artificial Intelligence Research}}  \bibinfo{volume}{74} (\bibinfo{year}{2022}).
\newblock


\bibitem[McCarthy et~al\mbox{.}(2004)]%
        {mccarthy2004dynamic}
\bibfield{author}{\bibinfo{person}{Kevin McCarthy}, \bibinfo{person}{James Reilly}, \bibinfo{person}{Lorraine McGinty}, {and} \bibinfo{person}{Barry Smyth}.} \bibinfo{year}{2004}\natexlab{}.
\newblock \showarticletitle{On the dynamic generation of compound critiques in conversational recommender systems}. In \bibinfo{booktitle}{\emph{Adaptive Hypermedia and Adaptive Web-Based Systems: Third International Conference, AH 2004, Eindhoven, The Netherlands, August 23-26, 2004. Proceedings 3}}. Springer, \bibinfo{pages}{176--184}.
\newblock


\bibitem[McHugh(2012)]%
        {mchugh2012interrater}
\bibfield{author}{\bibinfo{person}{Mary~L McHugh}.} \bibinfo{year}{2012}\natexlab{}.
\newblock \showarticletitle{Interrater reliability: the kappa statistic}.
\newblock \bibinfo{journal}{\emph{Biochemia medica}} \bibinfo{volume}{22}, \bibinfo{number}{3} (\bibinfo{year}{2012}), \bibinfo{pages}{276--282}.
\newblock


\bibitem[Miller(2019)]%
        {miller2019explanation}
\bibfield{author}{\bibinfo{person}{Tim Miller}.} \bibinfo{year}{2019}\natexlab{}.
\newblock \showarticletitle{Explanation in artificial intelligence: Insights from the social sciences}.
\newblock \bibinfo{journal}{\emph{Artificial intelligence}}  \bibinfo{volume}{267} (\bibinfo{year}{2019}), \bibinfo{pages}{1--38}.
\newblock


\bibitem[Moher et~al\mbox{.}(2009)]%
        {moher_preferred_2009}
\bibfield{author}{\bibinfo{person}{David Moher}, \bibinfo{person}{Alessandro Liberati}, \bibinfo{person}{Jennifer Tetzlaff}, {and} \bibinfo{person}{Douglas~G. Altman}.} \bibinfo{year}{2009}\natexlab{}.
\newblock \showarticletitle{Preferred reporting items for systematic reviews and meta-analyses: {The} {PRISMA} statement}.
\newblock \bibinfo{journal}{\emph{PLoS Medicine}} \bibinfo{volume}{6}, \bibinfo{number}{7} (\bibinfo{year}{2009}).
\newblock


\bibitem[Ouzzani et~al\mbox{.}(2016)]%
        {ouzzani2016rayyan}
\bibfield{author}{\bibinfo{person}{Mourad Ouzzani}, \bibinfo{person}{Hossam Hammady}, \bibinfo{person}{Zbys Fedorowicz}, {and} \bibinfo{person}{Ahmed Elmagarmid}.} \bibinfo{year}{2016}\natexlab{}.
\newblock \showarticletitle{Rayyan—a web and mobile app for systematic reviews}.
\newblock \bibinfo{journal}{\emph{Systematic reviews}}  \bibinfo{volume}{5} (\bibinfo{year}{2016}), \bibinfo{pages}{1--10}.
\newblock


\bibitem[Pecune et~al\mbox{.}(2019a)]%
        {pecune_model_2019}
\bibfield{author}{\bibinfo{person}{Florian Pecune}, \bibinfo{person}{Shruti Murali}, \bibinfo{person}{Vivian Tsai}, \bibinfo{person}{Yoichi Matsuyama}, {and} \bibinfo{person}{Justine Cassell}.} \bibinfo{year}{2019}\natexlab{a}.
\newblock \showarticletitle{A {Model} of {Social} {Explanations} for a {Conversational} {Movie} {Recommendation} {System}}. In \bibinfo{booktitle}{\emph{Proceedings of the 7th {International} {Conference} on {Human}-{Agent} {Interaction}}}. \bibinfo{publisher}{ACM}, \bibinfo{address}{Kyoto Japan}, \bibinfo{pages}{135--143}.
\newblock
\showISBNx{978-1-4503-6922-0}
\href{https://doi.org/10.1145/3349537.3351899}{doi:\nolinkurl{10.1145/3349537.3351899}}


\bibitem[Pecune et~al\mbox{.}(2019b)]%
        {pecune_exploratory_2019}
\bibfield{author}{\bibinfo{person}{Florian Pecune}, \bibinfo{person}{Magalie Ochs}, {and} \bibinfo{person}{Catherine Pelachaud}.} \bibinfo{year}{2019}\natexlab{b}.
\newblock \showarticletitle{An exploratory study of user experience in a sentiment-aware conversational agent}.
\newblock \bibinfo{journal}{\emph{International Journal of Human-Computer Studies}}  \bibinfo{volume}{131} (\bibinfo{year}{2019}).
\newblock


\bibitem[Pramod and Bafna(2022)]%
        {pramod2022conversational}
\bibfield{author}{\bibinfo{person}{Dhanya Pramod} {and} \bibinfo{person}{Prafulla Bafna}.} \bibinfo{year}{2022}\natexlab{}.
\newblock \showarticletitle{Conversational recommender systems techniques, tools, acceptance, and adoption: a state of the art review}.
\newblock \bibinfo{journal}{\emph{Expert Systems with Applications}}  \bibinfo{volume}{203} (\bibinfo{year}{2022}), \bibinfo{pages}{117539}.
\newblock


\bibitem[Purington et~al\mbox{.}(2017)]%
        {purington2017alexa}
\bibfield{author}{\bibinfo{person}{Amanda Purington}, \bibinfo{person}{Jessie~G Taft}, \bibinfo{person}{Shruti Sannon}, \bibinfo{person}{Natalya~N Bazarova}, {and} \bibinfo{person}{Samuel~Hardman Taylor}.} \bibinfo{year}{2017}\natexlab{}.
\newblock \showarticletitle{" Alexa is my new BFF" social roles, user satisfaction, and personification of the Amazon Echo}. In \bibinfo{booktitle}{\emph{Proceedings of the 2017 CHI conference extended abstracts on human factors in computing systems}}. \bibinfo{pages}{2853--2859}.
\newblock


\bibitem[Rana et~al\mbox{.}(2024)]%
        {rana2024user}
\bibfield{author}{\bibinfo{person}{Arpit Rana}, \bibinfo{person}{Scott Sanner}, \bibinfo{person}{Mohamed~Reda Bouadjenek}, \bibinfo{person}{Ronald Di~Carlantonio}, {and} \bibinfo{person}{Gary Farmaner}.} \bibinfo{year}{2024}\natexlab{}.
\newblock \showarticletitle{User experience and the role of personalization in critiquing-based conversational recommendation}.
\newblock \bibinfo{journal}{\emph{ACM Transactions on the Web}} \bibinfo{volume}{18}, \bibinfo{number}{4} (\bibinfo{year}{2024}), \bibinfo{pages}{1--21}.
\newblock


\bibitem[Samagaio et~al\mbox{.}(2021)]%
        {samagaio_chatbot_2021}
\bibfield{author}{\bibinfo{person}{Rui Samagaio}, \bibinfo{person}{David Carneiro}, {and} \bibinfo{person}{Paulo Novais}.} \bibinfo{year}{2021}\natexlab{}.
\newblock \showarticletitle{A chatbot for recipe recommendation and preference modeling}. In \bibinfo{booktitle}{\emph{Proceedings of the 19th {IEEE} {International} {Conference} on {Smart} {Technologies}}}. \bibinfo{pages}{130--137}.
\newblock


\bibitem[Schrepp et~al\mbox{.}(2017)]%
        {schrepp_design_2017}
\bibfield{author}{\bibinfo{person}{Martin Schrepp}, \bibinfo{person}{Andreas Hinderks}, {and} \bibinfo{person}{Jörg Thomaschewski}.} \bibinfo{year}{2017}\natexlab{}.
\newblock \showarticletitle{Design and evaluation of a short version of the user experience questionnaire ({UEQ}-{S})}.
\newblock \bibinfo{journal}{\emph{International Journal of Interactive Multimedia and Artificial Intelligence}} \bibinfo{volume}{4}, \bibinfo{number}{6} (\bibinfo{year}{2017}), \bibinfo{pages}{103--108}.
\newblock


\bibitem[Silva et~al\mbox{.}(2022)]%
        {silva_polite_2022}
\bibfield{author}{\bibinfo{person}{Tiago Silva}, \bibinfo{person}{Diana Gonçalves}, {and} \bibinfo{person}{Nelson Guimarães}.} \bibinfo{year}{2022}\natexlab{}.
\newblock \showarticletitle{Polite task-oriented dialog agent for enhancing user experience}. In \bibinfo{booktitle}{\emph{Proceedings of the 21st {ACM} {International} {Conference} on {Multimodal} {Interaction}}}. \bibinfo{pages}{243--252}.
\newblock


\bibitem[Siro et~al\mbox{.}(2023)]%
        {siro2023understanding}
\bibfield{author}{\bibinfo{person}{Clemencia Siro}, \bibinfo{person}{Mohammad Aliannejadi}, {and} \bibinfo{person}{Maarten De~Rijke}.} \bibinfo{year}{2023}\natexlab{}.
\newblock \showarticletitle{Understanding and predicting user satisfaction with conversational recommender systems}.
\newblock \bibinfo{journal}{\emph{ACM Transactions on Information Systems}} \bibinfo{volume}{42}, \bibinfo{number}{2} (\bibinfo{year}{2023}), \bibinfo{pages}{1--37}.
\newblock


\bibitem[Thom et~al\mbox{.}(2024)]%
        {thom2024nutria}
\bibfield{author}{\bibinfo{person}{Damian Thom}, \bibinfo{person}{Jefferson Ortega}, {and} \bibinfo{person}{Rosa Felix}.} \bibinfo{year}{2024}\natexlab{}.
\newblock \showarticletitle{NutrIA: An Out-of-the-Standard Nutritional Recommendation Mobile Application Powered by Artificial Intelligence}. In \bibinfo{booktitle}{\emph{2024 IEEE 4th International Conference on Advanced Learning Technologies on Education \& Research (ICALTER)}}. IEEE, \bibinfo{pages}{1--4}.
\newblock


\bibitem[Tintarev and Masthoff(2015)]%
        {tintarev2015explaining}
\bibfield{author}{\bibinfo{person}{Nava Tintarev} {and} \bibinfo{person}{Judith Masthoff}.} \bibinfo{year}{2015}\natexlab{}.
\newblock \showarticletitle{Explaining recommendations: Design and evaluation}.
\newblock In \bibinfo{booktitle}{\emph{Recommender systems handbook}}. \bibinfo{publisher}{Springer}, \bibinfo{pages}{353--382}.
\newblock


\bibitem[Tsai and Brusilovsky(2021)]%
        {tsai2021effects}
\bibfield{author}{\bibinfo{person}{Chun-Hua Tsai} {and} \bibinfo{person}{Peter Brusilovsky}.} \bibinfo{year}{2021}\natexlab{}.
\newblock \showarticletitle{The effects of controllability and explainability in a social recommender system}.
\newblock \bibinfo{journal}{\emph{User Modeling and User-Adapted Interaction}}  \bibinfo{volume}{31} (\bibinfo{year}{2021}), \bibinfo{pages}{591--627}.
\newblock


\bibitem[Verhagen et~al\mbox{.}(2024)]%
        {verhagen2024influence}
\bibfield{author}{\bibinfo{person}{Ruben~S Verhagen}, \bibinfo{person}{Alexandra Marcu}, \bibinfo{person}{Mark~A Neerincx}, {and} \bibinfo{person}{Myrthe~L Tielman}.} \bibinfo{year}{2024}\natexlab{}.
\newblock \showarticletitle{The Influence of Interdependence on Trust Calibration in Human-Machine Teams}.
\newblock In \bibinfo{booktitle}{\emph{HHAI 2024: Hybrid Human AI Systems for the Social Good}}. \bibinfo{publisher}{IOS Press}, \bibinfo{pages}{300--314}.
\newblock


\bibitem[Wang et~al\mbox{.}(2023)]%
        {wang2023rethinking}
\bibfield{author}{\bibinfo{person}{Qi Wang}, \bibinfo{person}{Yujian Zhu}, \bibinfo{person}{Yu Wang}, \bibinfo{person}{Hua Cheng}, \bibinfo{person}{Weinan Xu}, {and} \bibinfo{person}{Jing Li}.} \bibinfo{year}{2023}\natexlab{}.
\newblock \showarticletitle{Rethinking the Evaluation for Conversational Recommender Systems}.
\newblock \bibinfo{journal}{\emph{arXiv preprint arXiv:2305.10793}} (\bibinfo{year}{2023}).
\newblock


\bibitem[Wu et~al\mbox{.}(2024)]%
        {wu2024survey}
\bibfield{author}{\bibinfo{person}{Likang Wu}, \bibinfo{person}{Zhi Zheng}, \bibinfo{person}{Zhaopeng Qiu}, \bibinfo{person}{Hao Wang}, \bibinfo{person}{Hongchao Gu}, \bibinfo{person}{Tingjia Shen}, \bibinfo{person}{Chuan Qin}, \bibinfo{person}{Chen Zhu}, \bibinfo{person}{Hengshu Zhu}, \bibinfo{person}{Qi Liu}, {et~al\mbox{.}}} \bibinfo{year}{2024}\natexlab{}.
\newblock \showarticletitle{A survey on large language models for recommendation}.
\newblock \bibinfo{journal}{\emph{World Wide Web}} \bibinfo{volume}{27}, \bibinfo{number}{5} (\bibinfo{year}{2024}), \bibinfo{pages}{60}.
\newblock


\bibitem[Yun and Lim(2025)]%
        {yun2025user}
\bibfield{author}{\bibinfo{person}{Sojeong Yun} {and} \bibinfo{person}{Youn-kyung Lim}.} \bibinfo{year}{2025}\natexlab{}.
\newblock \showarticletitle{User Experience with LLM-powered Conversational Recommendation Systems: A Case of Music Recommendation}. In \bibinfo{booktitle}{\emph{Proceedings of the 2025 CHI Conference on Human Factors in Computing Systems}}. \bibinfo{pages}{1--15}.
\newblock


\bibitem[Zaidi et~al\mbox{.}(2024)]%
        {zaidi2024review}
\bibfield{author}{\bibinfo{person}{Subiya Zaidi}, \bibinfo{person}{Sawan Rai}, {and} \bibinfo{person}{Kapil Juneja}.} \bibinfo{year}{2024}\natexlab{}.
\newblock \showarticletitle{A Review of Existing Conversational Recommendation Systems}. In \bibinfo{booktitle}{\emph{2024 2nd International Conference on Disruptive Technologies (ICDT)}}. IEEE, \bibinfo{pages}{22--26}.
\newblock


\bibitem[Zhang et~al\mbox{.}(2024)]%
        {zhang2024navigating}
\bibfield{author}{\bibinfo{person}{Yizhe Zhang}, \bibinfo{person}{Yucheng Jin}, \bibinfo{person}{Li Chen}, {and} \bibinfo{person}{Ting Yang}.} \bibinfo{year}{2024}\natexlab{}.
\newblock \showarticletitle{Navigating user experience of chatgpt-based conversational recommender systems: The effects of prompt guidance and recommendation domain}.
\newblock \bibinfo{journal}{\emph{arXiv preprint arXiv:2405.13560}} (\bibinfo{year}{2024}).
\newblock


\bibitem[Zhao et~al\mbox{.}(2024)]%
        {zhao2024recommender}
\bibfield{author}{\bibinfo{person}{Zihuai Zhao}, \bibinfo{person}{Wenqi Fan}, \bibinfo{person}{Jiatong Li}, \bibinfo{person}{Yunqing Liu}, \bibinfo{person}{Xiaowei Mei}, \bibinfo{person}{Yiqi Wang}, \bibinfo{person}{Zhen Wen}, \bibinfo{person}{Fei Wang}, \bibinfo{person}{Xiangyu Zhao}, \bibinfo{person}{Jiliang Tang}, {et~al\mbox{.}}} \bibinfo{year}{2024}\natexlab{}.
\newblock \showarticletitle{Recommender systems in the era of large language models (llms)}.
\newblock \bibinfo{journal}{\emph{IEEE Transactions on Knowledge and Data Engineering}} (\bibinfo{year}{2024}).
\newblock


\bibitem[Zhou et~al\mbox{.}(2020)]%
        {zhou2020design}
\bibfield{author}{\bibinfo{person}{Li Zhou}, \bibinfo{person}{Jianfeng Gao}, \bibinfo{person}{Di Li}, {and} \bibinfo{person}{Heung-Yeung Shum}.} \bibinfo{year}{2020}\natexlab{}.
\newblock \showarticletitle{The design and implementation of xiaoice, an empathetic social chatbot}.
\newblock \bibinfo{journal}{\emph{Computational Linguistics}} \bibinfo{volume}{46}, \bibinfo{number}{1} (\bibinfo{year}{2020}), \bibinfo{pages}{53--93}.
\newblock


\end{thebibliography}





\end{document}